\documentclass[submission, Phys]{SciPost}
\begin{document}

\begin{center}{\Large \textbf{
Quantum Walks on Graphs of the Ordered Hamming Scheme and Spin Networks
}}\end{center}

% TODO: write the author list here. Use initials + surname format.
% Separate subsequent authors by a comma, omit comma at the end of the list.
% Mark the corresponding author with a superscript *.
\begin{center}
Hiroshi Miki\textsuperscript{1*},
Satoshi Tsujimoto\textsuperscript{2},
Luc Vinet\textsuperscript{3}
\end{center}

\begin{center}
{\bf 1} Meteorological College, Asahi-Cho, Kashiwa 277 0852, Japan
\\
{\bf 2} Department of Applied Mathematics and Physics, Graduate School of Informatics, 
Kyoto University, Sakyo-Ku, Kyoto 606 8501, Japan
\\
{\bf 3} Centre de Recherches Math\'{e}matiques, Universit\'{e} de Montr\'{e}al, PO Box 6128, 
Centre-ville Station, Montr\'{e}al (Qu\'{e}bec), H3C 3J7, Canada
\\
* hmiki@mc-jma.go.jp
\end{center}

\begin{center}
\today
\end{center}

% For convenience during refereeing: line numbers
%\linenumbers

\section*{Abstract}
{\bf
% TODO: write your abstract here.

It is shown that the hopping of a single excitation on certain triangular spin lattices with non-uniform couplings and local magnetic fields can be described as the projections of quantum walks on graphs of the ordered Hamming scheme of depth 2. For some values of the parameters the models exhibit perfect state transfer between two summits of the lattice. Fractional revival is also observed in some instances. The bivariate Krawtchouk polynomials of the Tratnik type that form the eigenvalue matrices of the ordered Hamming scheme of depth 2 give the overlaps between the energy eigenstates and the occupational basis vectors.
}

\vspace{10pt}
\noindent\rule{\textwidth}{1pt}
\tableofcontents\thispagestyle{fancy}
\noindent\rule{\textwidth}{1pt}
\vspace{10pt}

\section{Introduction}
\label{sec:intro}

This paper introduces two-dimensional spin lattices that exhibit perfect state transfer between two single locations and multi-site fractional revival on a one-dimensional subset of the lattice. These novel models are obtained by projecting quantum walks on graphs that belong to the ordered Hamming scheme which generalizes the well known Hamming association scheme.
On the one hand, continuous walks on graphs have been used to formulate various computation algorithms \cite{Childs2,Childs1,Farhi1}. On the other hand, the 1-excitation dynamics of spin chains has attracted attention as a mean to realize the transport of quantum states with a minimum of external controls \cite{Bose1,Bosse1,Kay1,Nikolopoulos1}. One speaks of perfect state transfer (PST) when the transport between two locations happens with probability one. It has been appreciated that non-uniform couplings and possibly local magnetic fields are required to achieve PST over distances of more than three sites \cite{Christandl1}. One analytic model that admits PST over (reasonably) arbitrary distances has couplings given by the Krawtchouk polynomials recurrence coefficients \cite{Albanese1}. Interestingly, it has been observed \cite{Christandl1} that the 1-excitation dynamics of this Krawtchouk chain can actually be obtained by projecting quantum walks on the 1-link hypercube to a weighted path. It will be recalled that the hypercube is one of the simplest graphs of the Hamming scheme \cite{Brouwer1,Stanton1}. The fact that the Krawtchouk polynomials arise naturally in that scheme is not foreign to the connection we just mentioned. The end-to-end PST in the chain can thus be seen as a manifestation of the fact that there is also PST between antipodal points of the hypercube. These results have motivated extensive examinations of quantum walks and especially of PST on graphs \cite{Godsil1,Kendon1}.

The coherent transport of states on higher dimensional spin lattices has also been explored. A few models with interesting transfer properties \cite{Miki2,Miki1,Post1} have been designed using the theory of multivariate Krawtchouk polynomials \cite{Diaconis1,Grunbaum1,Hoare1,Tratnik1}. These systems exhibit fractional revival (FR) whereby an initially localized state is reproduced periodically in a number of fixed locations \cite{Banchi1,Genest1}. In view of the relation between the hypercube and the Krawtchouk chain, it is natural to enquire if such models could not be obtained from the projection of quantum walks in higher dimensional graphs. In pursuing that question we will in fact identify graphs in a generalization of the Hamming scheme with dynamics that projects to 1-excitation hopping on a triangular lattice exhibiting perfect state transfer and multi-site fractional revival. We suggest that these systems could be realized as photonic lattices and possibly be of use for certain algorithms.

The paper will be organized as follows. The definition of the ordered Hamming scheme of depth 2 will be recalled in section 2. Particular graphs in that scheme will be identified in section 3 and the dynamics governed by their adjacency matrices will be shown to project to 1-excitation Hamiltonians for a triangular lattice of spins in the plane. The bivariate Krawtchouk polynomials of the Tratnik type will be introduced in section 4 to obtain the energy eigenstates. The transport properties will be examined in section 5 and it will be found that there is perfect state transfer between two specific summits of the triangular lattice. The paper will end with concluding remarks.

\section{The ordered Hamming scheme of depth 2}

Let $Q = {\mathbb Z}/2{\mathbb Z}$. 
Consider the set $Q^{(n,r)}$ of vectors of dimension $nr$ over $Q$. 
The vector $x \in Q^{(n,r)}$ will be presented by
the $r$-binary sequences of length $n$ over $Q$:
\begin{align*}
 {x}=(\overline x_{1}, \overline x_{2},\ldots, \overline x_{n}),
\end{align*}
where $\overline x_{j} = (x_{j1},x_{j2},\ldots ,x_{jr}) \in Q^{r}$.
We define the shape $e$ of $x \in Q^{(n,r)}$ by
\begin{align*}
&e(x)=(e_1,e_2,\cdots e_r),\\
&e_i=\# \{ j\in \{ 1,2,\cdots ,n\} ~|~x_{ji}=1,x_{j,i+1}=x_{j,i+2}=\cdots=x_{jr}=0\}
\end{align*}
and denote the set of the all shapes by 
\begin{align*}
E=e(Q^{(n,r)})=\{ (e_1,e_2,\cdots ,e_r) \in ({\mathbb Z}_{\ge 0})^n \mid 0\le e_1+e_2+\cdots +e_r\le n\}. 
\end{align*}
For example, $a=(00,10,11,00,01) \in Q^{(5,2)}$ is a $2$-binary sequence of length 5 and
$e(a)=(1,2)$.  
For two vectors $x, y \in Q^{(n,r)}$, we shall write $ x \sim_e y$ if the shape of $x-y$ is equal to $e$. 
Then we can introduce the graph $G_{e}$ associated with the shape $e$ as the one where all two vertices $(v_{x}, v_{y})$ in $\{v_{x} \mid x \in Q^{(n,r)}\}$ are linked if $v_{x} \sim_e v_{y}$; the corresponding adjacency matrix is given by
\[
A_{e}=(a_{x,y}),\quad a_{x,y}=\left\{
\begin{array}{cl}
 1 &  (x \sim_e y)\\
 0 &  (\textrm{otherwise})
\end{array}\right.
\]
It is known that ${\cal A}=\{ A_{e}~|~e\in E\}$ forms an association scheme. It is called the ordered Hamming scheme of depth $r$\cite{Bierbrauer,MartinStinson}. 
\par
In this paper, for fixed positive integer $N$, we shall consider the ordered Hamming scheme of depth 2 $(Q^{(N,2)},{\cal A})$ where the set of adjacency matrices
\[
A_{(i,j)}\quad 0\le i+j\le N
\]
form the (commutative) Bose-Mesner algebra:
\[
A_{(i,j)}A_{(k,l)}= \sum_{0\le i'+j'\le N} \alpha_{(i,j),(k,l)}^{(i',j')} A_{(i',j')}.
\]
The intersection numbers $\alpha_{(i,j),(k,l)}^{(i',j')}$ are equal to the number of vertices $z$ such that $e(x-z)=(i,j)$ and $e(y-z)=(k,l)$ if $e(x-y)=(i',j')$ for $x,y,z \in Q^{(N,2)}$.
In particular, one has the following explicit formulas involving $A_{(1,0)}$ and $A_{(0,1)}$.
\begin{align}
A_{(1,0)}A_{(i,j)}&=(N+1-i-j)A_{(i-1,j)}+jA_{(i,j)}+(i+1)A_{(i+1,j)} \label{bose-mesner1}\\
A_{(0,1)}A_{(i,j)}&=2(N+1-i-j)A_{(i,j-1)}+2(i+1)A_{(i+1,j-1)} \label{bose-mesner2}\\ 
&+(j+1)A_{(i-1,j+1)}+(j+1)A_{(i,j+1)}. \nonumber
\end{align}
It is not difficult to obtain the above relations.
Let us write 
\[
A_{(1,0)}A_{(i,j)}=a_{i,j}A_{(i-1,j)}+b_{i,j}A_{(i,j)}+c_{i,j}A_{(i+1,j)}.
\]
The coefficient $a_{i,j}$ stands for how many $z$ exist such that $e(x-z)={(1,0)}$ and {$e(y-z)=(i,j)$} if $e(x-y)=(i-1,j)$. 
For example, take ${y}=(00,00,\cdots ,00)$ and ${x}=(10,10,\cdots ,10,01,01,\cdots ,01, 00,00,\cdots ,00)$ with $i-1$ $10$s, $j$ $01$s ($11$s or the combination of $01$ and $11$ is also possible) and $N+1-i-j$ $00$s.
In this situation, $z$ must be obtained by changing one of the $N+1-i-j$ $00$s by $10$ and there are $N+1-i-j$ ways of doing this.
It is easy to see that the number of ways does not depend on the choice of the elements $y\in Q^{(N,2)}$.
We thus have $a_{i,j}=N+1-i-j$.
The number $c_{i,j}$ of $z$ such that $e(x-z)={(1,0)}$ and {$e(y-z)=(i,j)$} if $e(x-y)=(i+1,j)$ can similarly be obtained.
Take ${y}=(00,00,\cdots ,00)$ and ${x}$ consisting of $i+1$ $10$s and $j$ $01$s and $N-1-i-j$ $00$s.
We see that a $z$ can be obtained by changing one of the $i+1$ $10$s by $00$s and there are $i+1$ ways of doing that. It thus follows that $c_{i,j}=i+1$.
For $b_{i,j}$, take ${y}=(00,00,\cdots ,00)$ and $x=(10,10,\cdots ,10,01,01,\cdots ,01,00,00,\cdots ,00)$ with $i$ $10$s, $j$ $01$s and $N-i-j$ $00$s.
Clearly a $z$ such that $e({x}-z)=(1,0)$ and $e({y}-z)=(i,j)$ if $e(x-y)=(i,j)$ can be obtained by changing one of the $j$ $01$s by $11$ and there are $j$ possible choices. As a result $b_{i,j}=j$ and \eqref{bose-mesner1} holds.\par
Formula \eqref{bose-mesner2} is derived in the same fashion. We shall therefore only indicate how the factor in front of $A_{(i,j-1)}$ is identified.
This coefficient stands for how many $z$ there are such that $e(x-z)=(0,1)$ and $e(y-z)=(i,j)$ if $e(x-y)=(i,j-1)$. Take $y=(00,00,\cdots ,00)$ and $x=(10,10,\cdots ,10,01,01,\cdots ,01, 00,00,\cdots ,00)$ with {$i$} $10$s, {$j-1$} $01$s and $N+1-i-j$ $00$s. To obtain such a $z$, we must change one of the $N+1-i-j$
$00$s by $01$ or $11$ and there are $2(N+1-i-j)$ ways of doing this.

\section{Special weighted graphs and their projections}
Let us consider the graph $G_{\alpha,\beta}$ whose adjacency matrix is $\alpha A_{(1,0)}+\beta A_{(0,1)}$ with $\alpha, \beta \in {\mathbb R}_{\ge 0}$. We shall call this graph ordered Hamming graph (of depth 2).
Now, following \cite{Bernard,Christandl1}, we consider the projection of the quantum walk on the ordered Hamming graph $G_{\alpha,\beta}$ to the ``column subspaces'' to identify the corresponding spin lattice.

To the vertices $ x\in V=Q^{(N,2)}$ ($|V|=(2^2)^{N}=4^{N}$), we associate orthonormalized vectors $\left| x\right> $ such that
\[
\langle\, x \mid y \,\rangle =
\begin{cases}
1 \quad (x\sim_{(0,0)} y)\\ 0\quad (\textrm{otherwise}) 
\end{cases}
\]
for $x,y \in V$.
In this notation the entries $A_{xy}$ of $A$ can be written as $\left< x | A| y \right>$.
Let $(0)\equiv (00,00,\cdots ,00)$ denote a corner and organize $V$ as a set of $\binom{N+2}{2}$ columns $V_{i,j}\quad (0\le i+j\le N)$ defined by
\[
V_{i,j}=\{ x\in V,\quad e(x)=(i,j)\}.
\]
The number $k_{i,j}$ of vertices in the column $V_{i,j}$ is given by
\[
k_{i,j}=\binom{N}{i,j}2^{j},
\]
where $\binom{N}{i,j}=\frac{N!}{i!j!(N-i-j)!}$ is the trinomial coefficient. The number $k_{i,j}$ can be identified by taking it into account that vertices with shape $(i,j)$ consist of $i$ $10$s, $j$ $01$ or $11$s and $N-i-j$ $00$s. 
Let us then label the vertices in column $V_{i,j}$ by $V_{(i,j),k},k=1,\dots k_{i,j}$.
Under the relation defined by the shape $(1,0)$, each $V_{(i,j),k}$ in the column $V_{i,j}$ is connected to $N-i-j$ elements of column $V_{i+1,j}$. Similarly, according to the association generated by the shape $(0,1)$, each $V_{(i,j),k}$ in $V_{i,j}$ is linked with $2(N-i-j)$ elements of column $V_{i,j+1}$ and $j$ elements of column $V_{i+1,j-1}$. Furthermore, with respect to $(1,0)$ each $V_{(i,j),k}$ in $V_{i,j}$ is connected to $j$ elements of the same column $V_{(i,j)}$.

Consider now the column space taken to be the linear span of the column
vectors given by 
\[
\left|\mathrm{col} ~i,j\right>=\frac{1}{\sqrt{k_{i,j}}}\sum_{k=1}^{k_{i,j}} \left| V_{(i,j),k}\right>,\quad 0\le i+j\le N.
\]
Since every vertex in
column $V_{i,j}$ is connected to the same number of vertices in columns $V_{i+1,j}$, $V_{i,j+1}$ and $V_{i+1,j-1}$ with respect to $(1,0)$ and $(0,1)$, $A_{(1,0)}$ and $A_{(0,1)}$ preserve the column space.  

Let us compute the matrix elements of $A_{(1,0)}$ and $A_{(0,1)}$ in the basis of the column subspace.
\begin{align*}
\left< \mathrm{col}~i+1,j\right| A_{(1,0)}\left|\mathrm{col} ~i,j\right> &= \frac{1}{\sqrt{k_{i+1,j}k_{i,j}}}\sum_{l=1}^{k_{i+1,j}}\sum_{k=1}^{k_{i,j}}\left< V_{(i+1,j),l}\right| A_{(1,0)} \left| V_{(i,j),k}\right> \\
&=\frac{1}{\sqrt{k_{i+1,j}k_{i,j}}}(N-i-j)k_{i,j}\\
&= \sqrt{(i+1)(N-i-j)}.
\end{align*}
To derive the second relation, one can first pick a vertex in $V_{i,j}$, compute the scalar products
with the $(N-i-j)$ vertices to which it is linked in $V_{i+1,j}$ and then sum
over the $k_{i,j}$ vertices in $V_{i,j}$.
Similary, we also have
\begin{align*}
\left< \mathrm{col}~i,j\right| A_{(1,0)}\left|\mathrm{col} ~i,j\right> &= j,\\
\left< \mathrm{col}~i,j+1\right| A_{(0,1)}\left|\mathrm{col} ~i,j\right> &= \sqrt{2(j+1)(N-i-j)},\\
\left< \mathrm{col}~i+1,j-1\right| A_{(0,1)}\left|\mathrm{col} ~i,j\right> &= \sqrt{2(i+1)j}.
\end{align*}
To conclude, the quantum walk on the ordered Hamming graph $G_{\alpha,\beta}$ is equivalent
to the one-excitation dynamics of the spin lattice of triangular shape governed by the following Hamiltonian:
\begin{align}
\begin{split}\label{hamiltonian}
H=&\sum_{0\le i+j\le N} 
\alpha \sqrt{(i+1)(N-i-j)}\frac{\sigma_{i,j}^x \sigma_{i+1,j}^x+\sigma_{i,j}^y\sigma_{i+1,j}^y}{2}\\
&\quad \quad +\beta \sqrt{2(j+1)(N-i-j)}\frac{\sigma_{i,j}^x \sigma_{i,j+1}^x+\sigma_{i,j}^y\sigma_{i,j+1}^y}{2}\\
&\quad \quad +\beta \sqrt{2(i+1)j}\frac{\sigma_{i,j}^x \sigma_{i+1,j-1}^x+\sigma_{i,j}^y\sigma_{i-1,j+1}^y}{2}+ \alpha j \frac{1+\sigma_{i,j}^z}{2}.
\end{split}
\end{align}
The lattice sites are labelled by two integers $i,j$ between $0$ and $N$ and such that their sum is smaller or equal to $N$.
Indeed, on the subspace spanned by the 1-excitation orthonormal basis vectors $\left| e_{i,j}\right)\,(0 \le i+j \le N)$, 
%with $E_{i,j}$ the $(N+1)\times (N+1)$ matrix with $1$ in the $(i,j)$ entry and zeros everywhere else, 
we see that
\begin{align}\label{hamiltonian-action}
\begin{split}
H\left| e_{i,j}\right) &= \alpha \sqrt{(i+1)(N-i-j)}\left| e_{i+1,j}\right) + \beta \sqrt{2(j+1)(N-i-j)}\left| e_{i,j+1}\right)\\
&+ \alpha j\left| e_{i,j}\right) + \alpha \sqrt{i(N+1-i-j)}\left| e_{i-1,j}\right)  + \beta \sqrt{2j(N+1-i-j)}\left| e_{i,j-1}\right) \\
&+ \beta \sqrt{2(i+1)j}\left| e_{i+1,j-1}\right)+ \beta \sqrt{2i(j+1)} \left| e_{i-1,j+1}\right),
\end{split}
\end{align}
thus confirming the equivalence. 
%\textcolor{red}{The explanation for the notation $\left| \cdot \right) $ should be added?}

\section{Bivariate Krawtchouk polynomials and energy eigenstates}

In the Hamming scheme, the univariate Krawtchouk polynomials
\[
K_n^N(x;p)=~_{2}F_1 \left( 
\begin{array}{c}
-n, -x \\
-N
\end{array};\frac{1}{p}
\right)=\sum_{l=0}^{\infty }\frac{(-n)_l(-x)_l}{l!(-N)_l}\left( \frac{1}{p}\right)^l,\quad x,n=0,1,\cdots ,N
\]
come up and are applied to analyze the properties of the quantum walks on the associated graphs \cite{Christandl1}, where $(a)_n=a(a+1)\cdots (a+n-1)$ is the standard Pochhammer symbol. In the ordered Hamming scheme of depth 2, the bivariate Krawtchouk polynomials of the Tratnik type appear, as pointed out in the related coding theory \cite{Bierbrauer}. These two-variable orthogonal polynomials are defined as the following product of the univariate Krawtchouk polynomials:
\begin{eqnarray*}
T_{m,n}^N(x,y)= \frac{1}{(-N)_{m+n}}k_{m}^{N-n}(x;p)k_n^{N-x}(y;\frac{q}{1-p}),\quad 0\le x+y,m+n\le N,
\end{eqnarray*}
where $k_n^N(x;p)=(-N)_nK_n^N(x;p)$. The bivariate Krawtchouk polynomials are orthogonal with respect to the trinomial distribution function:
\[
\sum_{0\le x+y\le N} \binom{N}{x,y}p^x q^y(1-p-q)^{N-x-y}
T_{i,j}(x,y)T_{k,l}(x,y)
=\frac{(1-p-q)^{i+j}}{\binom{N}{i,j}\tilde{p}^i \tilde{q}^j}\delta_{i,k}\delta_{j,l},
\]
where $\tilde{p}=\frac{p(1-p-q)}{1-p},\tilde{q}=\frac{q}{1-p}$. These polynomials are also known to satisfy the 3-term recurrence relations involving multiplication by $x$ \begin{align}\label{tratnik1}
\begin{split}
xT_{i,j}^N(x,y)&=-p(N-i-j)(T_{i+1,j}^N(x,y)-T_{i,j}^N(x,y))\\
&-(1-p)i(T_{i-1,j}^N(x,y)-T_{i,j}^N(x,y))
\end{split}
\end{align}
and the 7-term recurrence relations when multiplied by $y$
\begin{align}\label{tratnik2}
\begin{split}
yT_{i,j}^N(x,y)&=\frac{pq}{1-p}(N-i-j)(T_{i+1,j}^N(x,y)-T_{i,j}^N(x,y))\\
&-\frac{q}{1-p}(N-i-j) (T_{i,j+1}^N(x,y)-T_{i,j}^N(x,y))\\
&+qi(T_{i-1,j}^N(x,y)-T_{i,j}^N(x,y))\\
&-(1-p-q)j(T_{i,j-1}^N(x,y)-T_{i,j}^N(x,y))\\
&-\frac{p(1-p-q)}{1-p}j(T_{i+1,j-1}^N(x,y)-T_{i,j}^N(x,y))\\
&-\frac{q}{1-p}i(T_{i-1,j+1}^N(x,y)-T_{i,j}^N(x,y)).
\end{split}
\end{align}
Furthermore, one has the generating function formula \cite{Genest2}:
\begin{align}\label{generating}
\sum_{0\le x+y\le N}\binom{N}{x,y}s^x t^y T_{i,j}^N(x,y)=(1+s+t)^{N-i-j}(1+\frac{p-1}{p}s+t)^i(1+\frac{p+q-1}{q}t)^j.
\end{align}\par
In the following, set 
\[
p=\frac{1}{2},\quad q=\frac{1}{4}
\]
and introduce the orthonormal bivariate Krawtchouk polynomials
\[
t_{i,j}^N(x,y) = \sqrt{\binom{N}{i,j}\tilde{p}^i \tilde{q}^j(1-p-q)^{-i-j}}T_{i,j}^N(x,y).
\]
From \eqref{tratnik1} and \eqref{tratnik2}, one can obtain for $\{ t_{i,j}^N(x,y)\}$ the following contiguity relation:
\begin{align}\label{tratnik-rec}
\begin{split}
\lambda _{x,y}t_{i,j}^N(x,y)&=\alpha \sqrt{(i+1)(N-i-j)}t_{i+1,j}^N(x,y)+\beta \sqrt{2(j+1)(N-i-j)}t_{i,j+1}^N(x,y)\\
           &+\alpha jt_{i,j}^N(x,y)+\alpha \sqrt{i(N+1-i-j)}t_{i-1,j}^N(x,y)\\
           &+\beta \sqrt{2j(N+1-i-j)}t_{i,j-1}^N(x,y)\\
           &+\beta \sqrt{2i(j+1)}t_{i-1,j+1}^N(x,y)+\beta \sqrt{2(i+1)j}t_{i+1,j-1}^N(x,y),
           \end{split}
           \end{align}
where the spectrum $\lambda_{x,y}$ is given by
\begin{equation}\label{spectrum}
\lambda_{x,y}=\alpha (N-2x)+ \beta (2N-2x-4y).
\end{equation}
It is a straightforward matter to identify the correspondance between the projection \eqref{hamiltonian-action} to the spin lattice of the quantum walk on the ordered Hamming graph $G_{\alpha ,\beta }$ and the above recurrence relations for bivariate Krawtchouk polynomials \eqref{tratnik-rec}.

\section{Transfer properties on the graphs}
Let us now examine the properties of the quantum walk on the ordered Hamming graph of depth 2 and of the projected dynamics on the spin lattice. With the motion initiated at $\left| e_{0,0}\right)$, the essential quantity is the transition amplitude 
\[
f_{(i,j)}(t)= \left( e_{i,j}| e^{-itH} | e_{0,0}\right).
\]\par
From the correspondence between \eqref{hamiltonian-action} and \eqref{tratnik-rec}, the Hamiltonian \eqref{hamiltonian} on 1-excitation subspace spanned by $\left|e_{i,j}\right)$ can be diagonalized by the bivariate Krawtchouk polynomials and its spectrum is given by \eqref{spectrum}.
With the overlaps between the $1$-excitation eigenstates of $H$ and the occupation basis states given by the orthonormalized polynomials $t_{i,j}^N(x,y)$ and using the generating function formula \eqref{generating}, one finds
\begin{align*}
f_{(i,j)}(t) & = \sum_{0\le x+y\le N} \binom{N}{x,y}\left( \frac{1}{2}\right)^x \left( \frac{1}{4}\right)^y \left( \frac{1}{4}\right)^{N-x-y} t_{0,0}^N(x,y)t_{(i,j)}^N(x,y)e^{-i\lambda_{x,y}t}\\
&= e^{-iN(\alpha +2\beta )t}\frac{\sqrt{2^j}}{4^N}\sqrt{\binom{N}{i,j}}(1+2z_1+z_2)^{N-i-j}(1-2z_1+z_2)^i(1-z_2)^{j},
\end{align*}
where $z_1=e^{2(\alpha +\beta )ti},z_2=e^{4\beta  ti}$. In \cite{Miki1,Post1}, fractional revival from the apex $(0,0)$ to the hypotenuse line $(i,j) \quad (i+j=N)$ was found in 2-dimensional $X\!X$-spin lattices related to the bivariate Krawtchouk polynomials of the Rahman type \cite{Diaconis1,Genest2,Grunbaum1,Hoare1}. To realize here a transfer to the same set or subset of points with $i+j=N$, it is easy to see that we should require that there be a time $t=T$ for which
\begin{equation} \label{cond1}
1+2z_1+z_2 = 0\quad (\exists T \in \mathbb{R}).
\end{equation}
Since $|z_1|=|z_2|=1$, the relation \eqref{cond1} simultaneously imposes that 
\begin{equation} \label{cond2}
z_2=1
\end{equation}
at the same time $T$. Quite interestingly, these instances are the conditions for perfect state transfer:
\[
|f_{(N,0)}(T)| =1,\quad |f_{(i,j)}(T)|=0\quad ((i,j)\ne (N,0)).
\]
Let us now clarify this. We can rewrite the condition \eqref{cond1} and \eqref{cond2}
as follows:
\[
e^{2i(\alpha +\beta )T }=-1,\quad e^{4i\beta T}=1,
\]
from where one finds
\[
(2\alpha T,2\beta T)= ((2m+1)\pi ,2n\pi ),(2m\pi ,(2n+1)\pi )\quad (m,n\in \mathbb{Z}).
\]
Therefore, we can conclude that if 
\[
\frac{\alpha }{\beta }= \frac{(\textrm{even integer })}{(\textrm{odd integer})}
\]
or
\[
\frac{\alpha }{\beta }= \frac{(\textrm{odd integer })}{(\textrm{even integer})},
\]
PST from $(0,0)$ to $(N,0)$ takes place at some time $T$.\par
\begin{figure}[htbp]
  \begin{center}
    \begin{tabular}{c}

      % 1
      \begin{minipage}{0.33\hsize}
        \begin{center}
          \includegraphics[width=3cm]{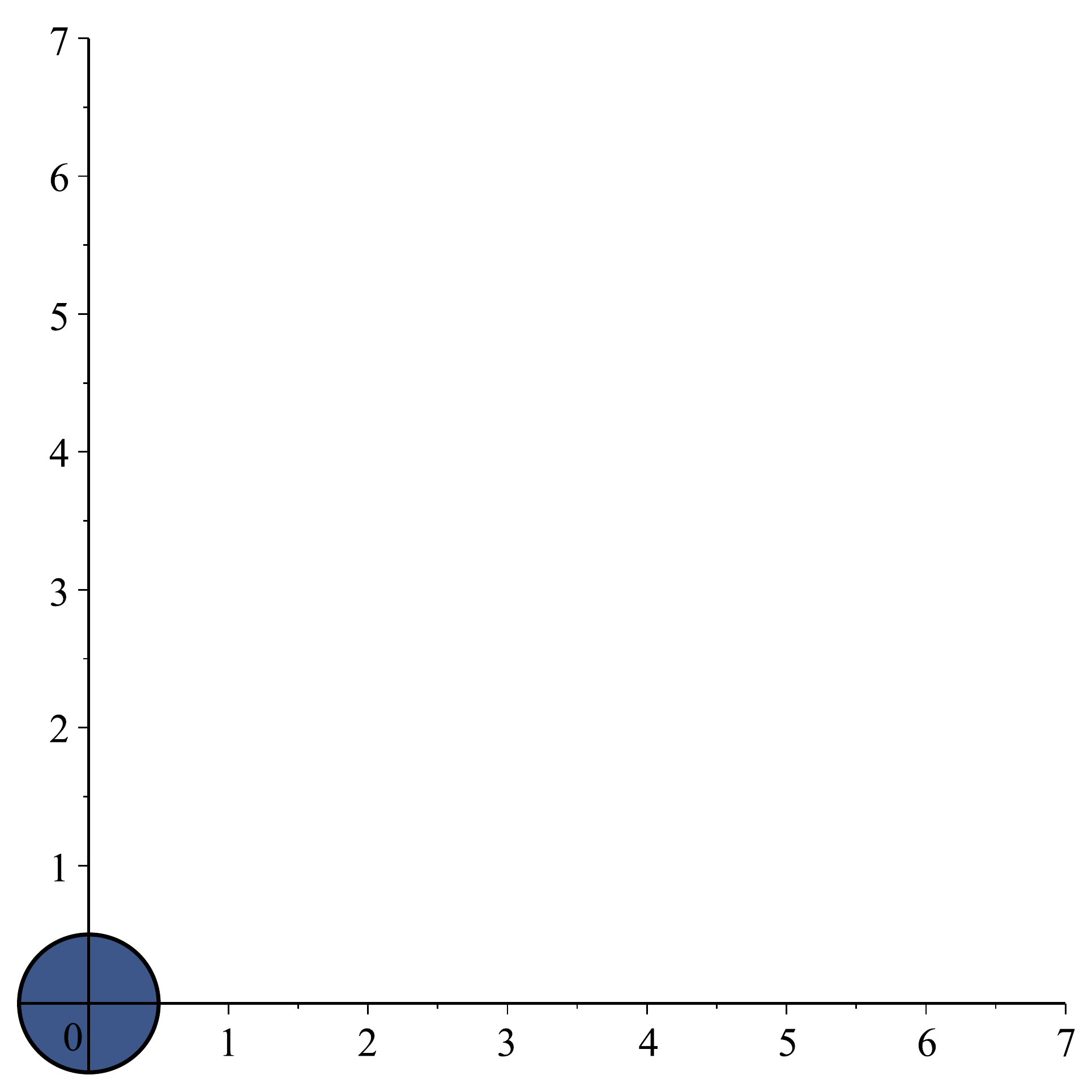}
         \hspace{1.6cm} $t=0$
        \end{center}
      \end{minipage}

      % 2
      \begin{minipage}{0.33\hsize}
        \begin{center}
          \includegraphics[width=3cm]{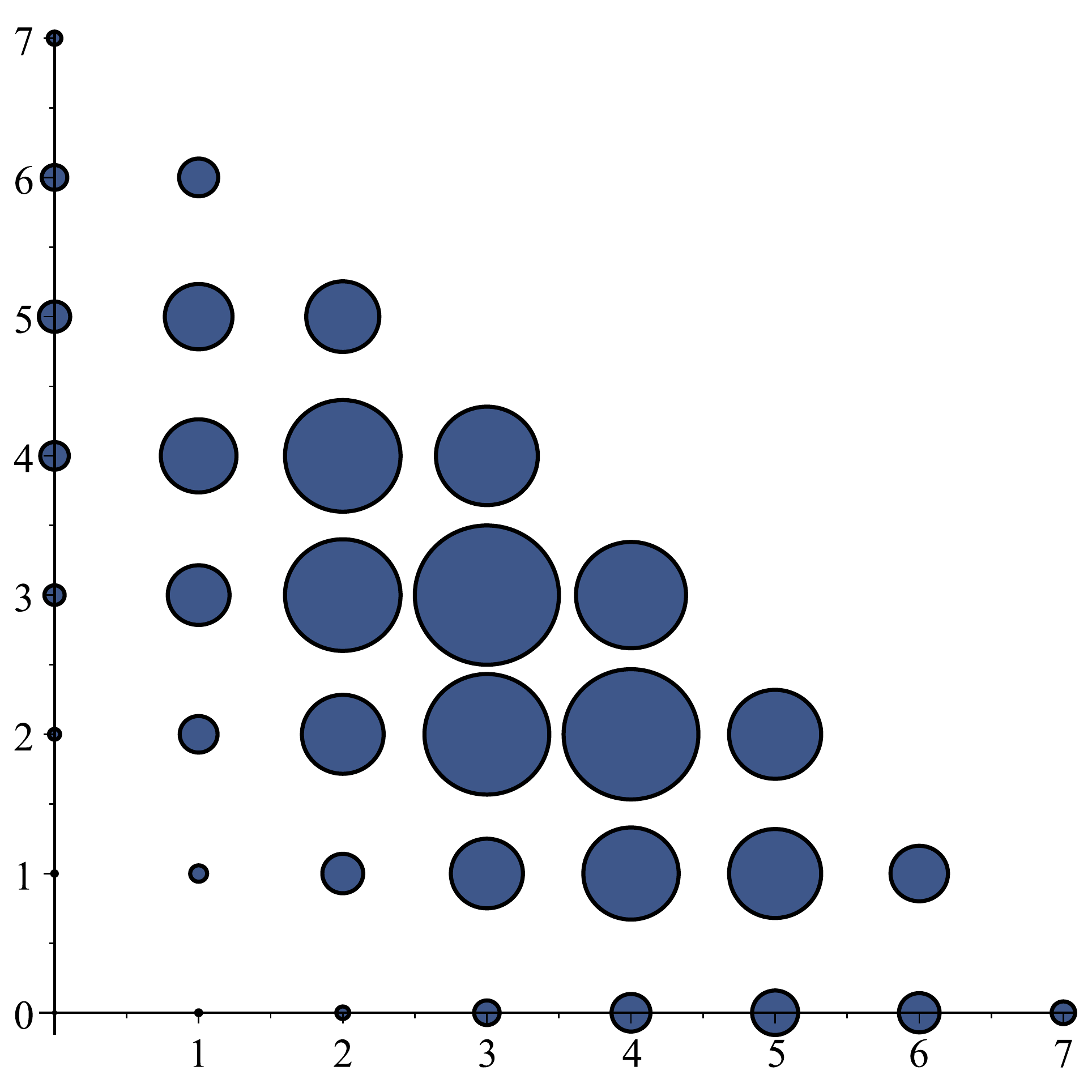}
                   \hspace{1.6cm} $t=\frac{\pi }{6}$
        \end{center}
      \end{minipage}

      % 3
      \begin{minipage}{0.33\hsize}
        \begin{center}
          \includegraphics[width=3cm]{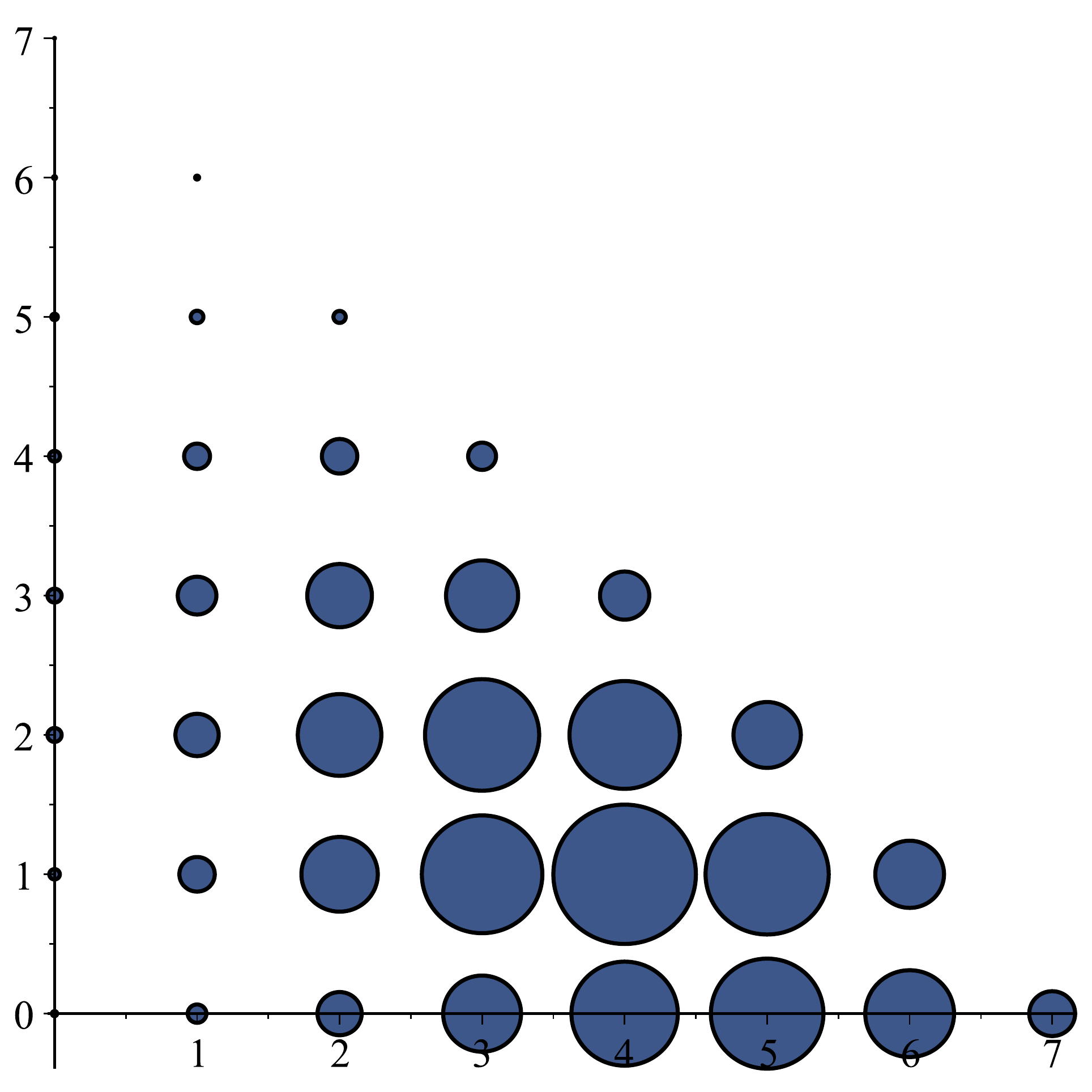}
                   \hspace{1.6cm} $t=\frac{\pi}{5}$
        \end{center}
      \end{minipage}
    \end{tabular}
        \begin{tabular}{c}

      % 1
      \begin{minipage}{0.33\hsize}
        \begin{center}
          \includegraphics[width=3cm]{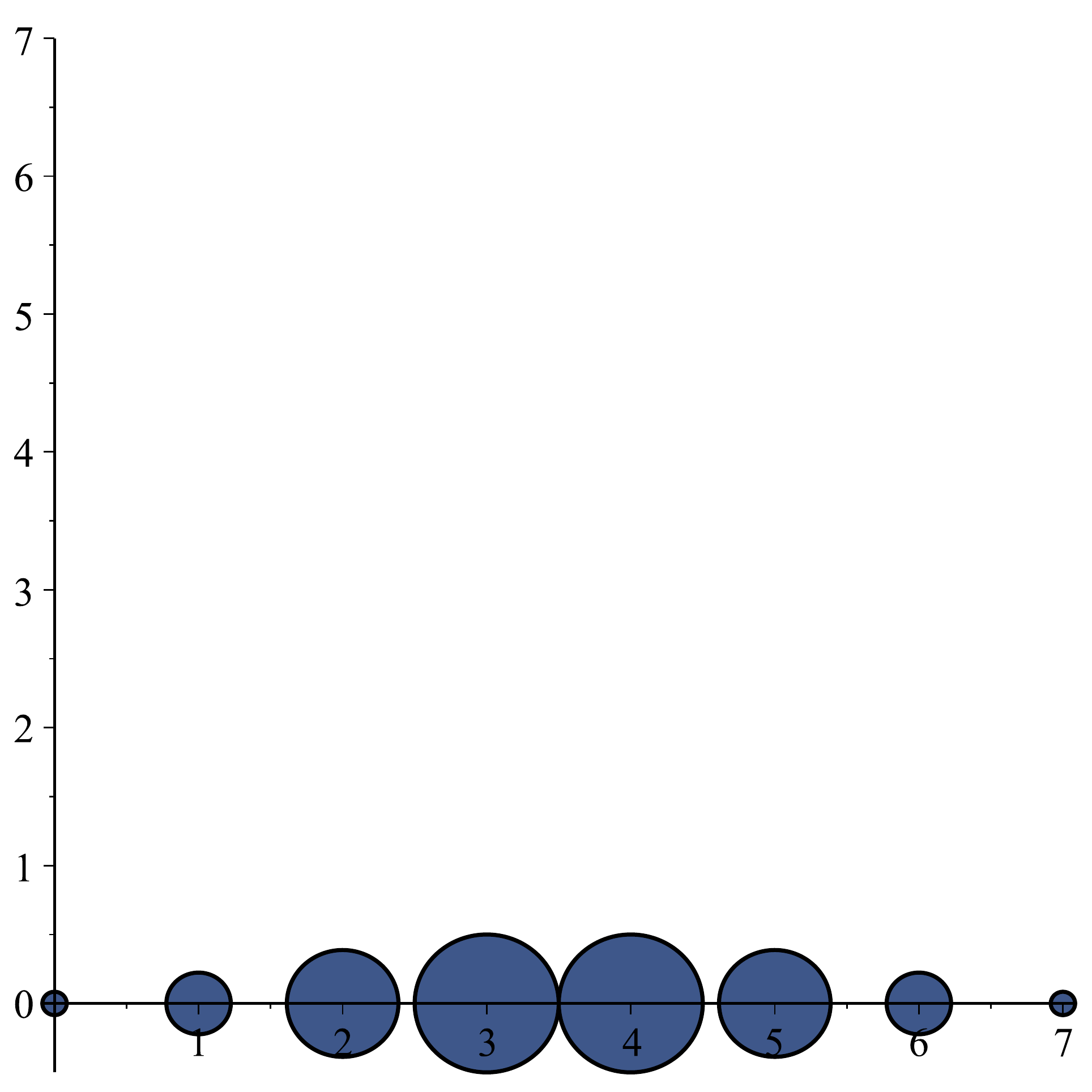}
                   \hspace{1.6cm} $t=\frac{\pi}{4}$
        \end{center}
      \end{minipage}

      % 2
      \begin{minipage}{0.33\hsize}
        \begin{center}
          \includegraphics[width=3cm]{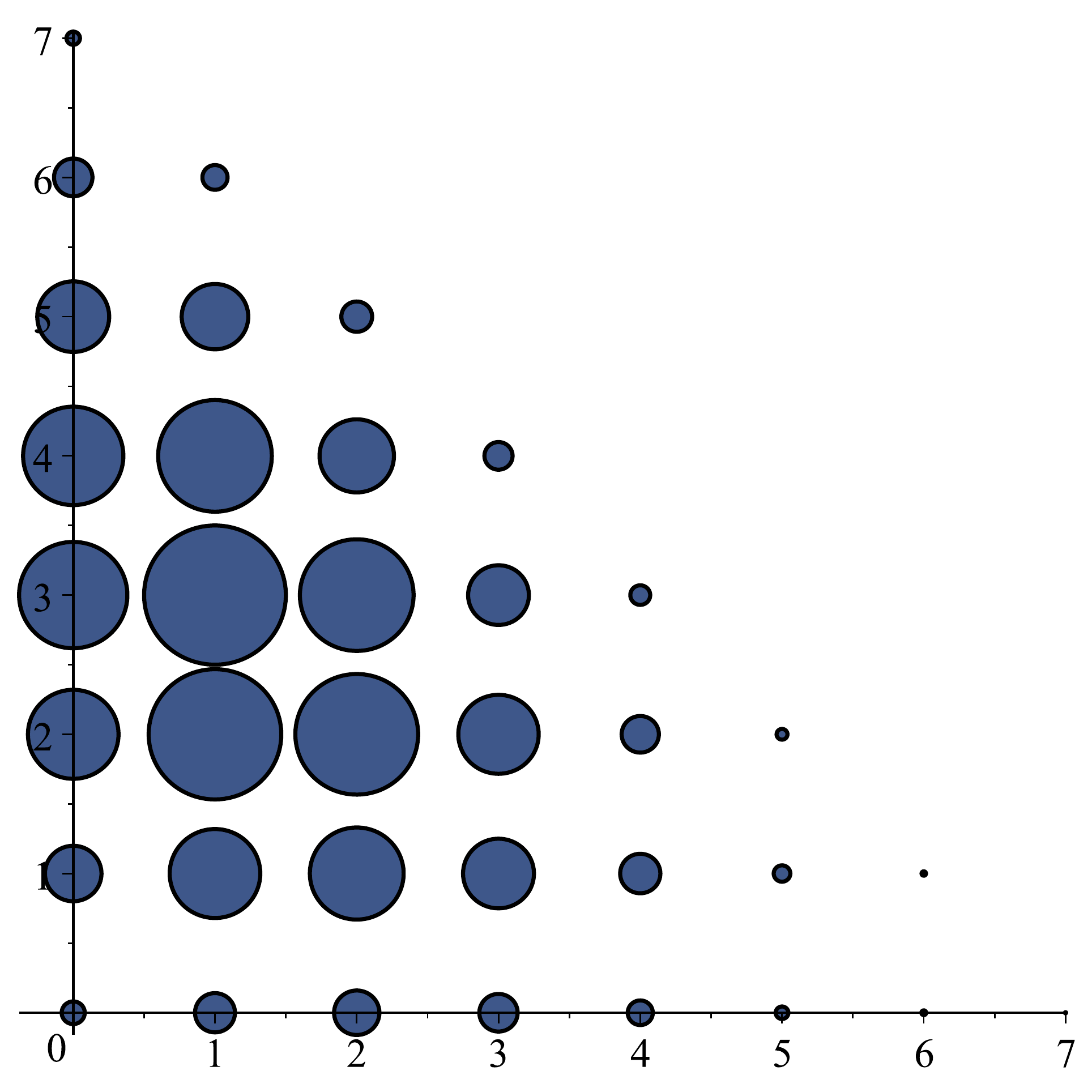}
                   \hspace{1.6cm} $t=\frac{\pi}{3}$
        \end{center}
      \end{minipage}

      % 3
      \begin{minipage}{0.3\hsize}
        \begin{center}
          \includegraphics[width=3cm,clip]{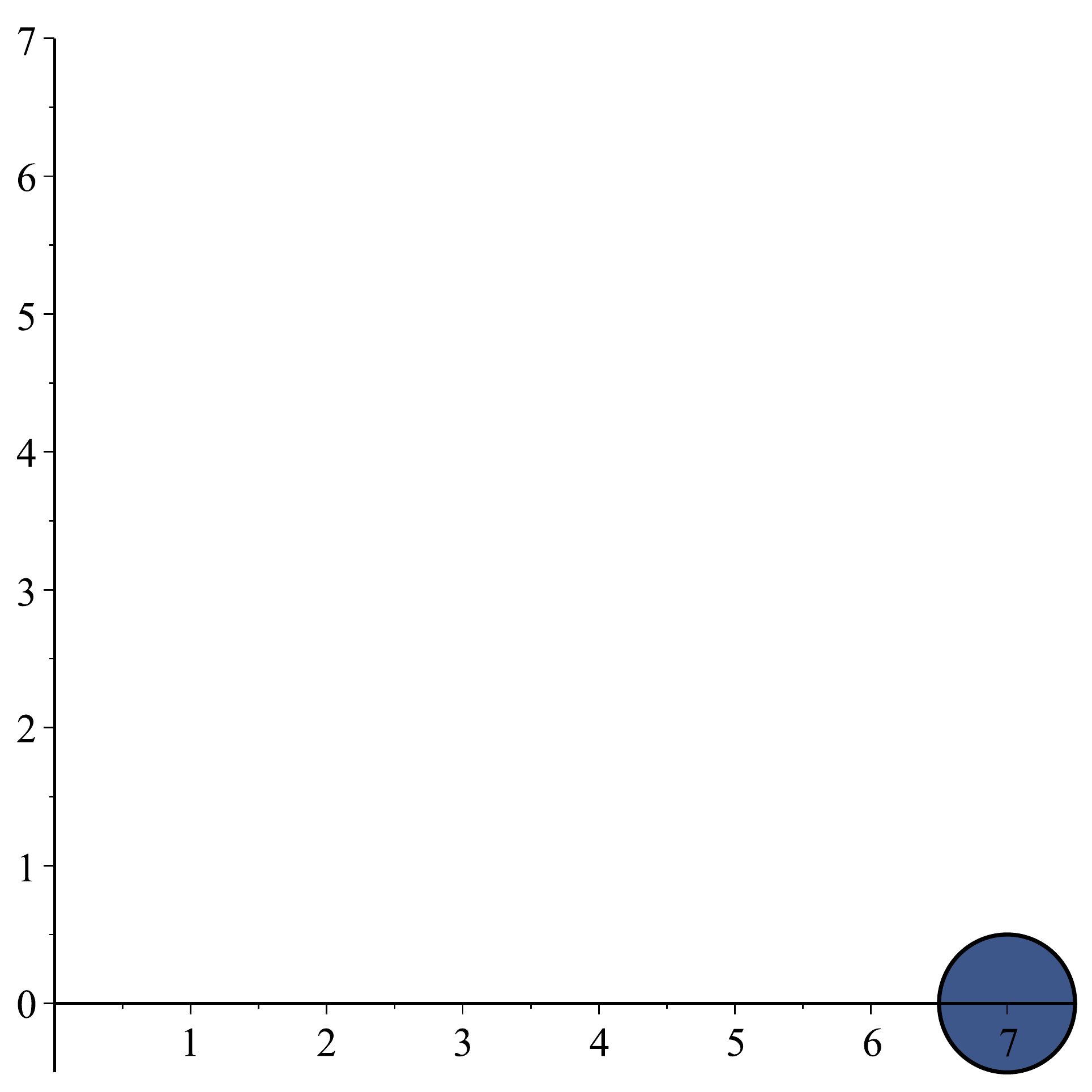}
        \end{center}
                 \hspace{1.6cm} $t=\frac{\pi}{2}$
      \end{minipage}
    \end{tabular}
    \caption{The transition amplitude $|f_{i,j}(t)|$ for $A_{(1,0)}+2A_{(0,1)}$ when $N=7$. The areas of the circles
are proportional to $|f_{(i,j)}(t)|$ at the given lattice point $(i,j)$. PST occurs at $\frac{\pi}{2}$ and FR on the set of sites $i=0,1,\cdots ,N$ and $j=0$ occurs at $t=\frac{\pi}{4}$.}
    \label{fig:a1b2}
  \end{center}
\end{figure}
\begin{figure}[htbp]
  \begin{center}
    \begin{tabular}{c}

      % 1
      \begin{minipage}{0.33\hsize}
        \begin{center}
          \includegraphics[width=3cm]{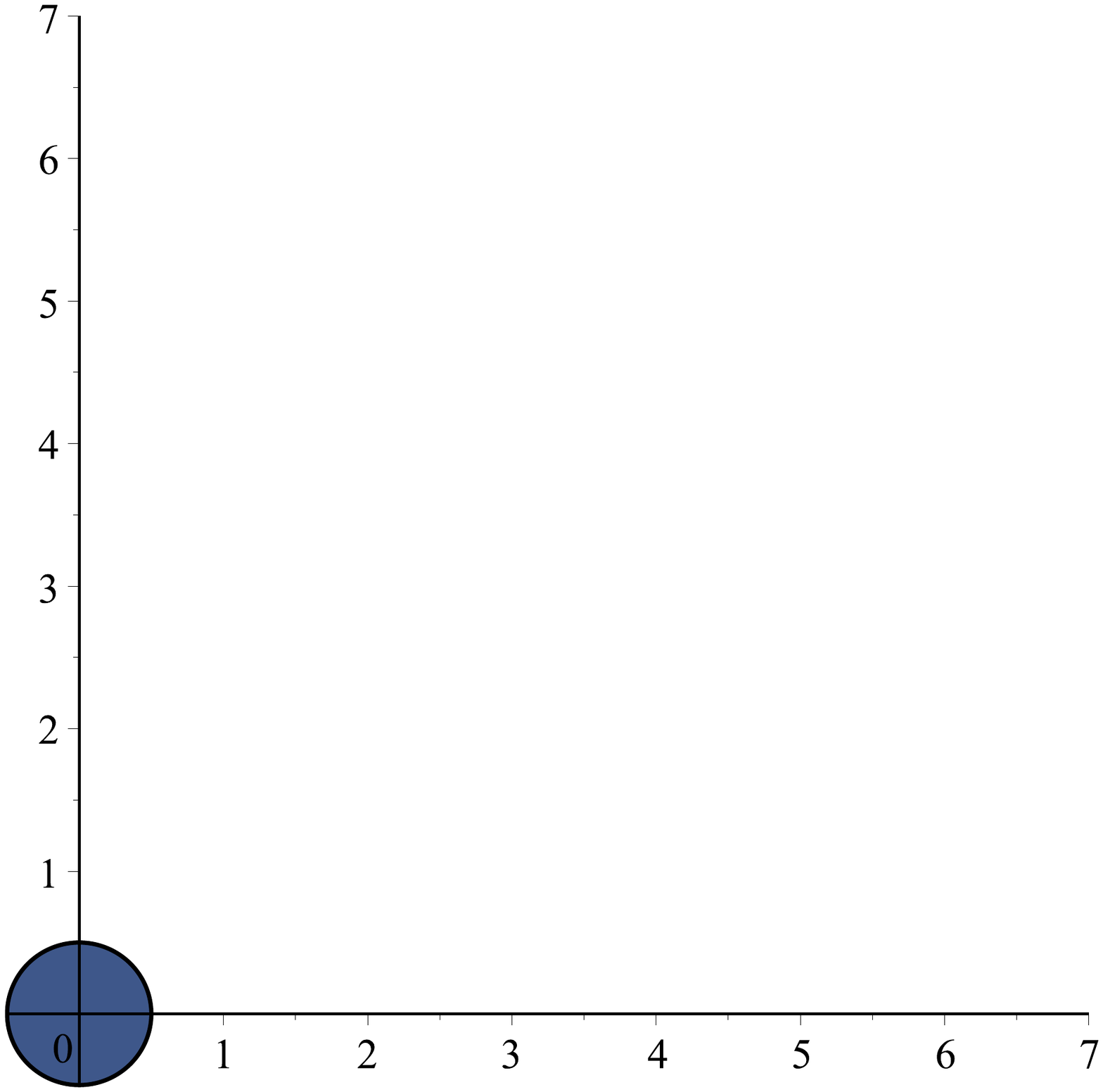}
         \hspace{1.6cm} $t=0$
        \end{center}
      \end{minipage}

      % 2
      \begin{minipage}{0.33\hsize}
        \begin{center}
          \includegraphics[width=3cm]{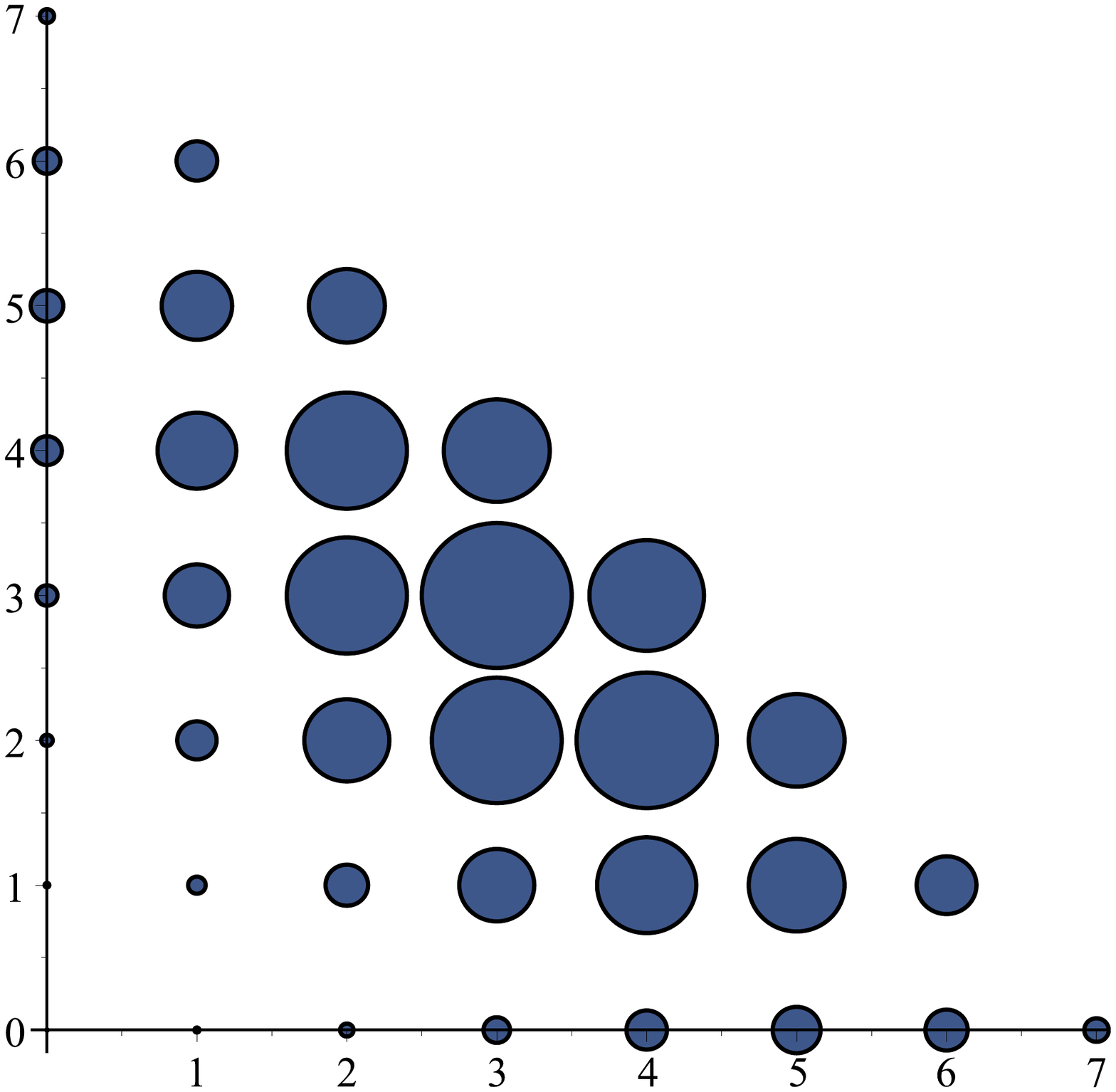}
                   \hspace{1.6cm} $t=\frac{\pi }{6}$
        \end{center}
      \end{minipage}

      % 3
      \begin{minipage}{0.33\hsize}
        \begin{center}
          \includegraphics[width=3cm]{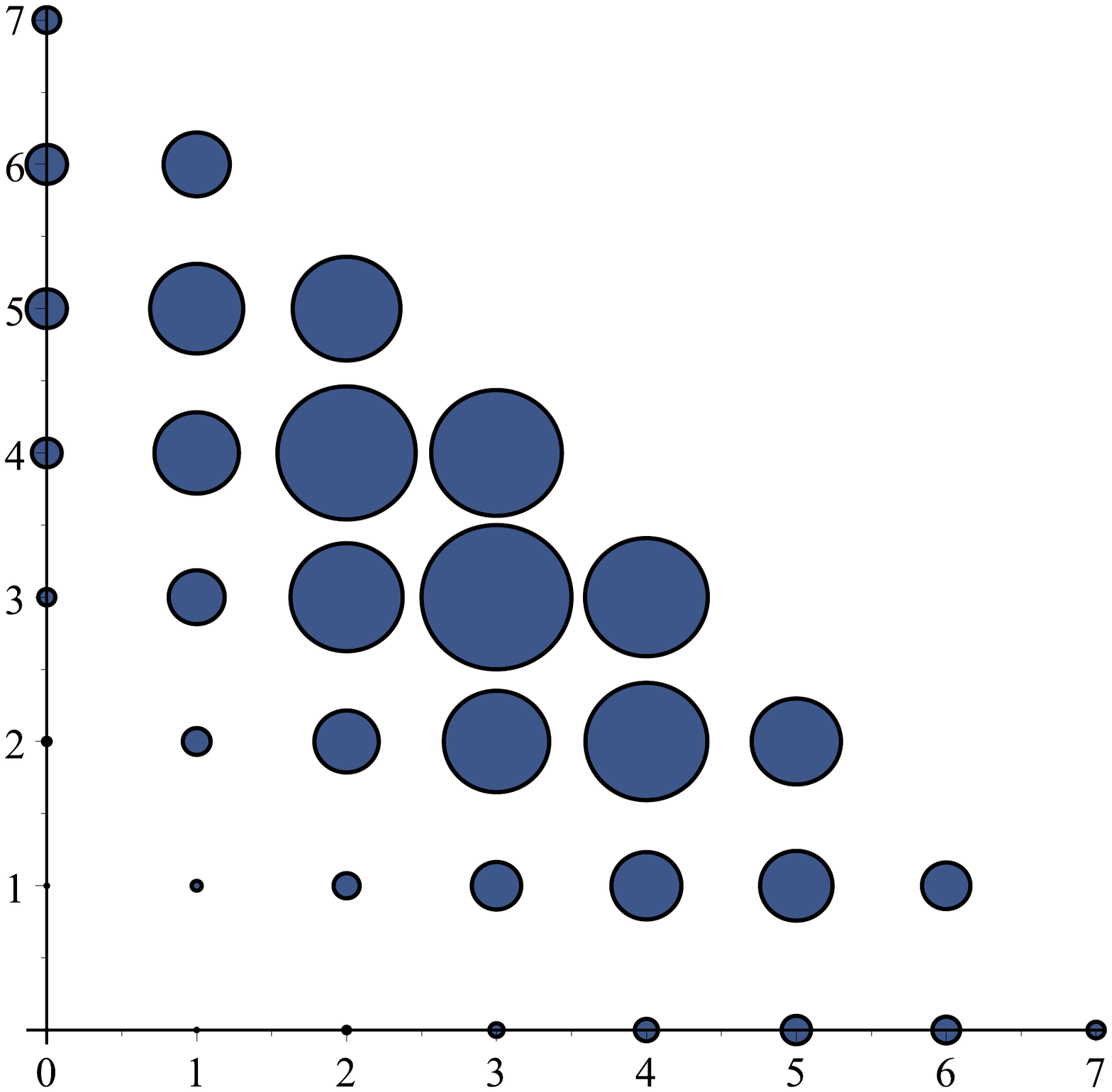}
                   \hspace{1.6cm} $t=\frac{\pi}{5}$
        \end{center}
      \end{minipage}
    \end{tabular}
        \begin{tabular}{c}

      % 1
      \begin{minipage}{0.33\hsize}
        \begin{center}
          \includegraphics[width=3cm]{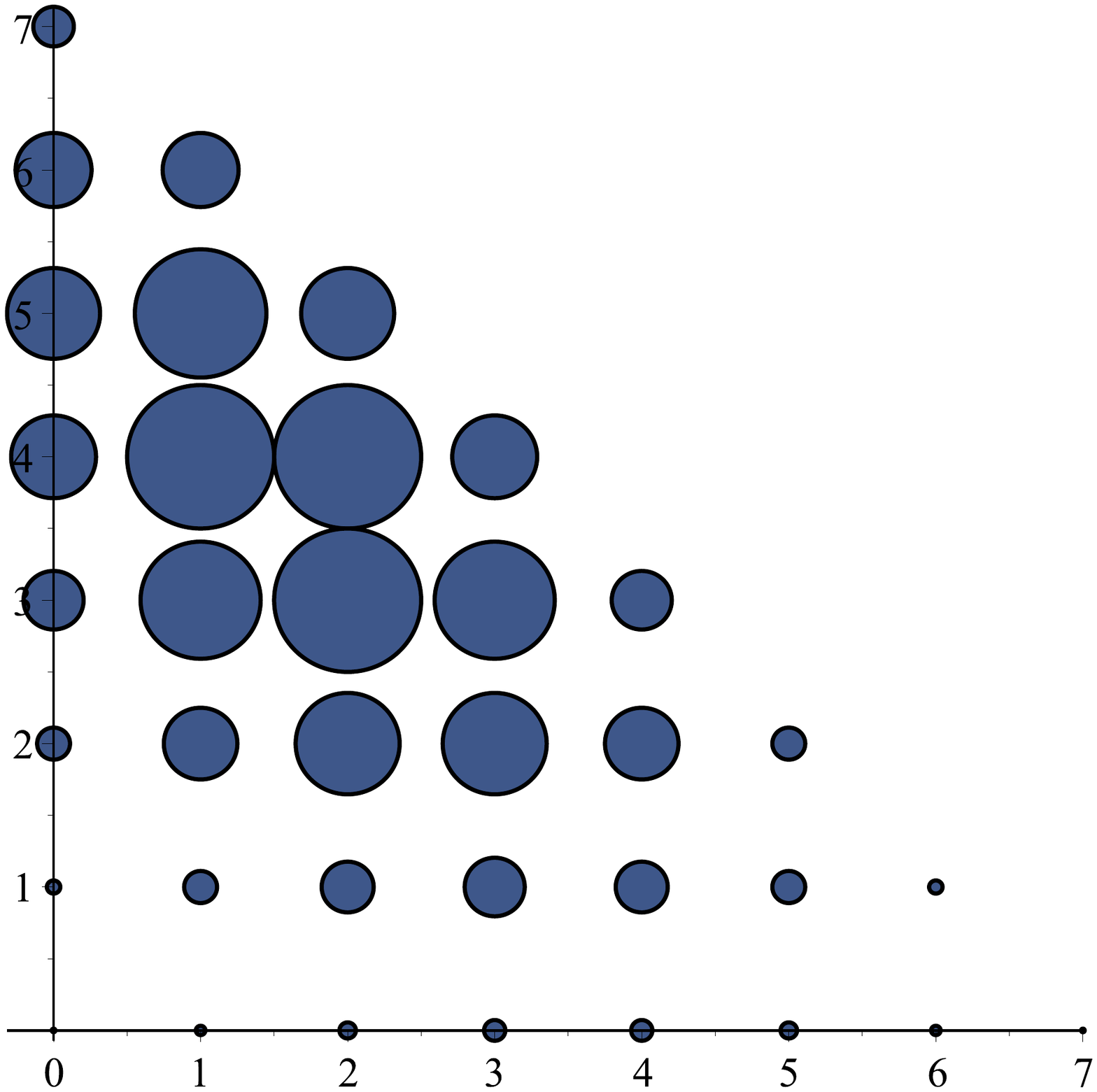}
                   \hspace{1.6cm} $t=\frac{\pi}{4}$
        \end{center}
      \end{minipage}

      % 2
      \begin{minipage}{0.33\hsize}
        \begin{center}
          \includegraphics[width=3cm]{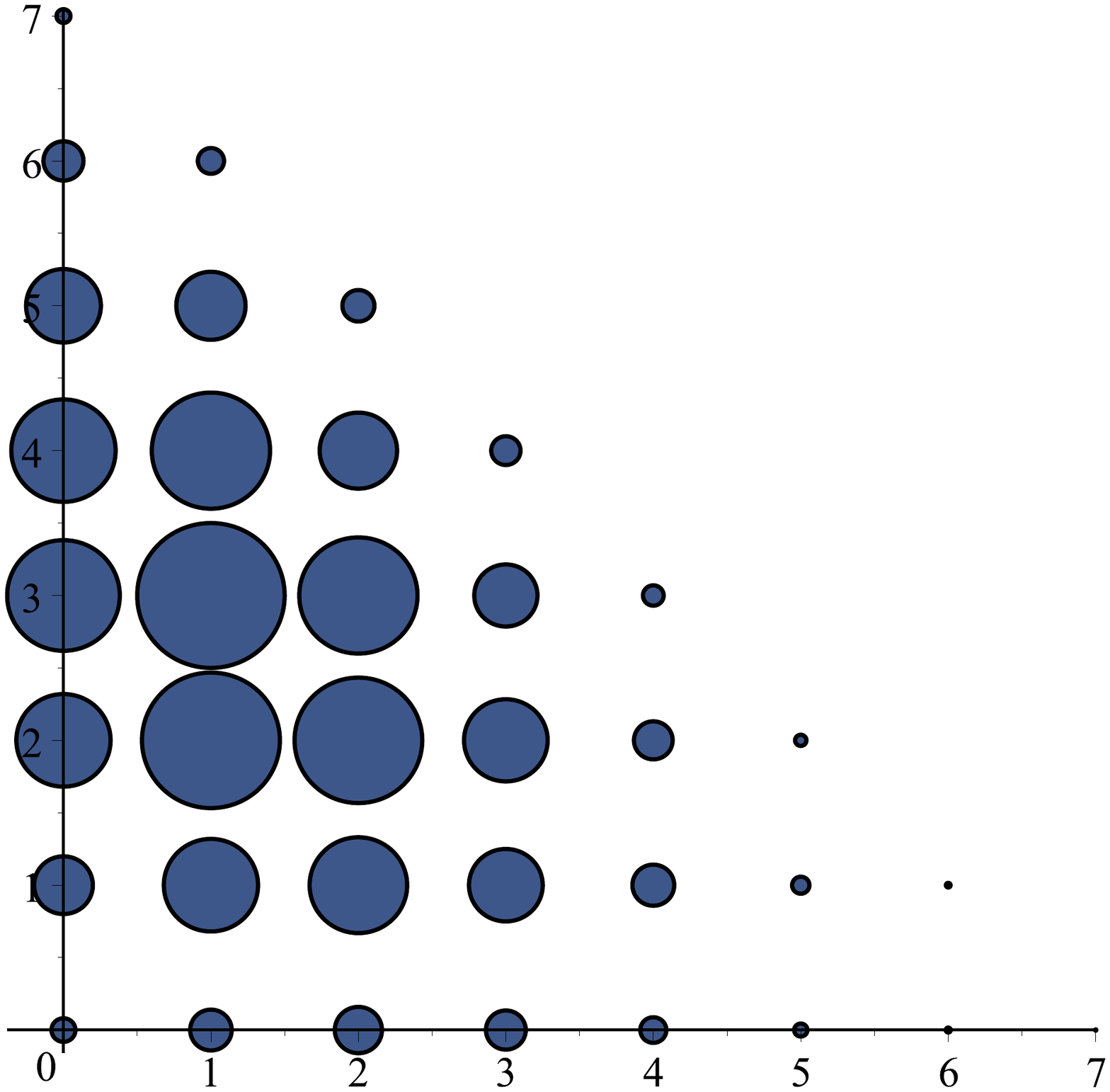}
                   \hspace{1.6cm} $t=\frac{\pi}{3}$
        \end{center}
      \end{minipage}

      % 3
      \begin{minipage}{0.33\hsize}
        \begin{center}
          \includegraphics[width=3cm]{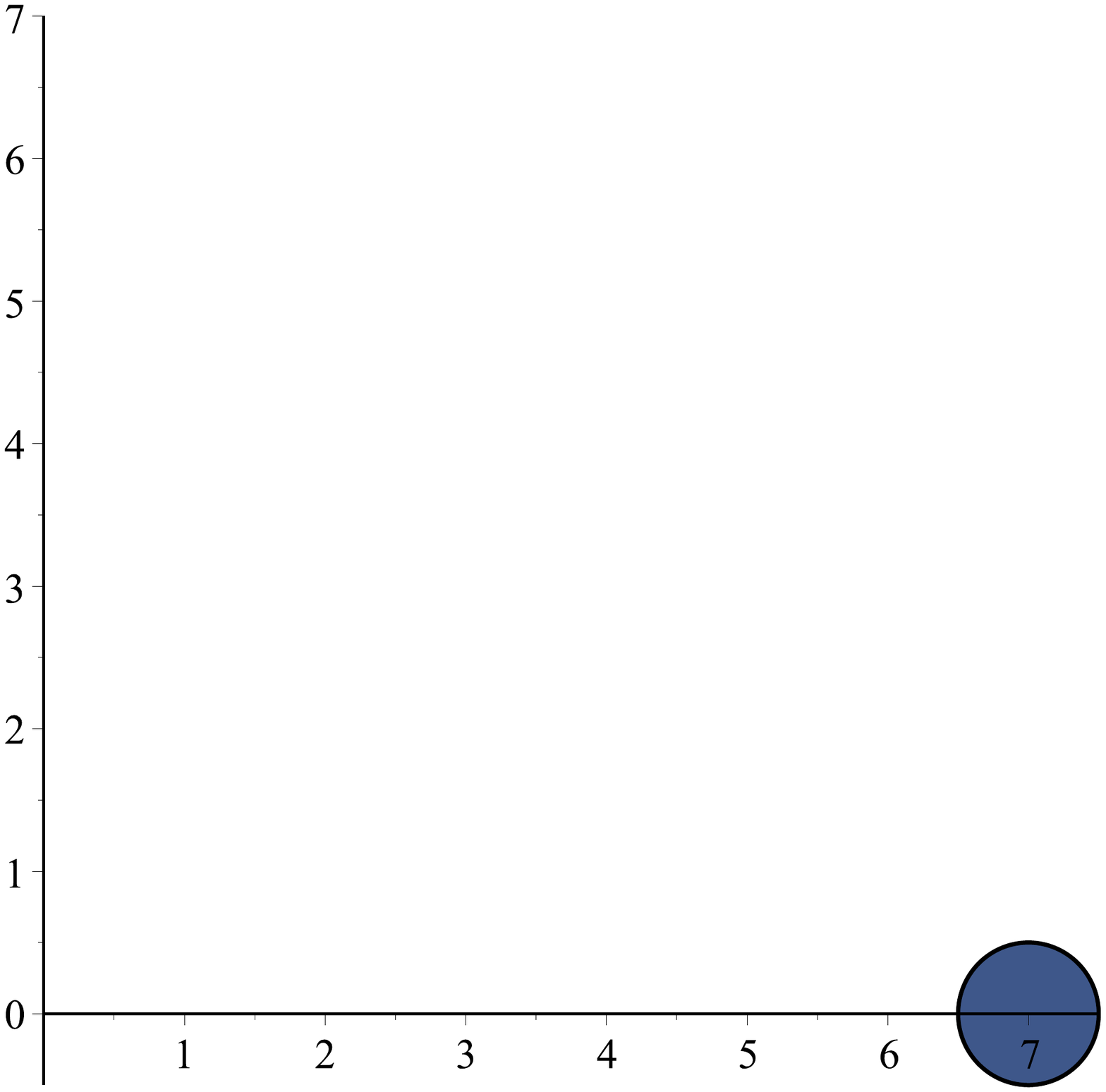}
        \end{center}
                 \hspace{1.6cm} $t=\frac{\pi}{2}$
      \end{minipage}
    \end{tabular}
    \caption{The transition amplitude $|f_{i,j}(t)|$ for $2A_{(1,0)}+A_{(0,1)}$ when $N=7$. The areas of the circles
are proportional to $|f_{(i,j)}(t)|$ at the given lattice point $(i,j)$. PST occurs at $\frac{\pi}{2}$.}
    \label{fig:a2b1}
  \end{center}
\end{figure}

The Fig.\ref{fig:a1b2} and Fig.\ref{fig:a2b1} are the plots of the transition probabilities of the graph $G_{\alpha,\beta}$ associated with $\alpha A_{(1,0)}+\beta A_{(0,1)}$.

\begin{figure}[htbp]

   \begin{minipage}{0.47\hsize}
        \begin{center}
          \includegraphics[width=4.5cm]{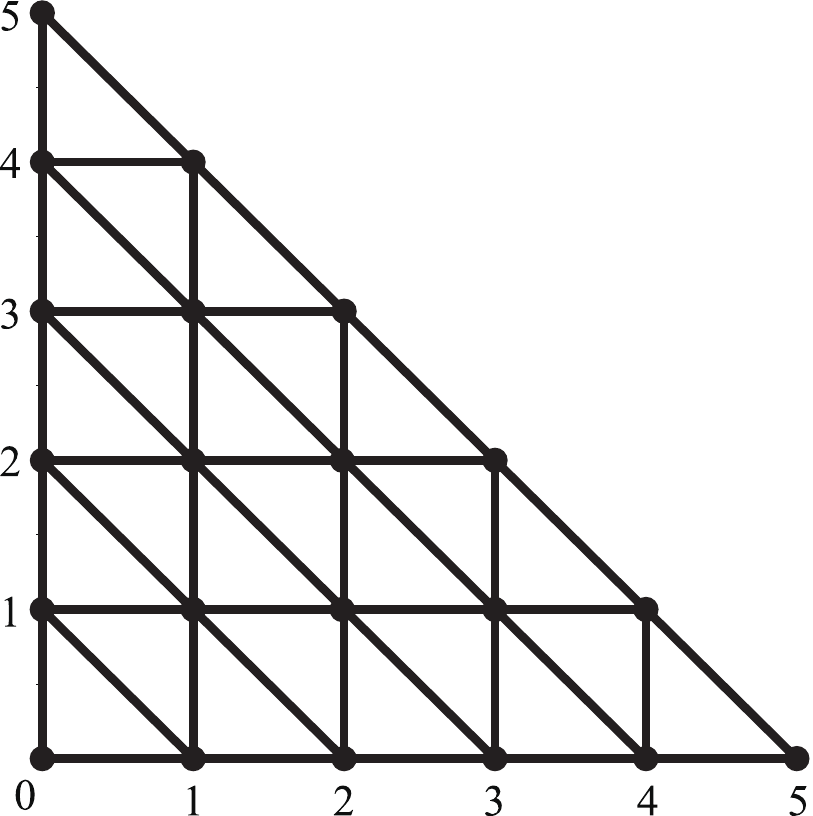}
        \end{center}
      \end{minipage}
      % 2
      \begin{minipage}{0.47\hsize}
        \begin{center}
          \includegraphics[width=4.5cm]{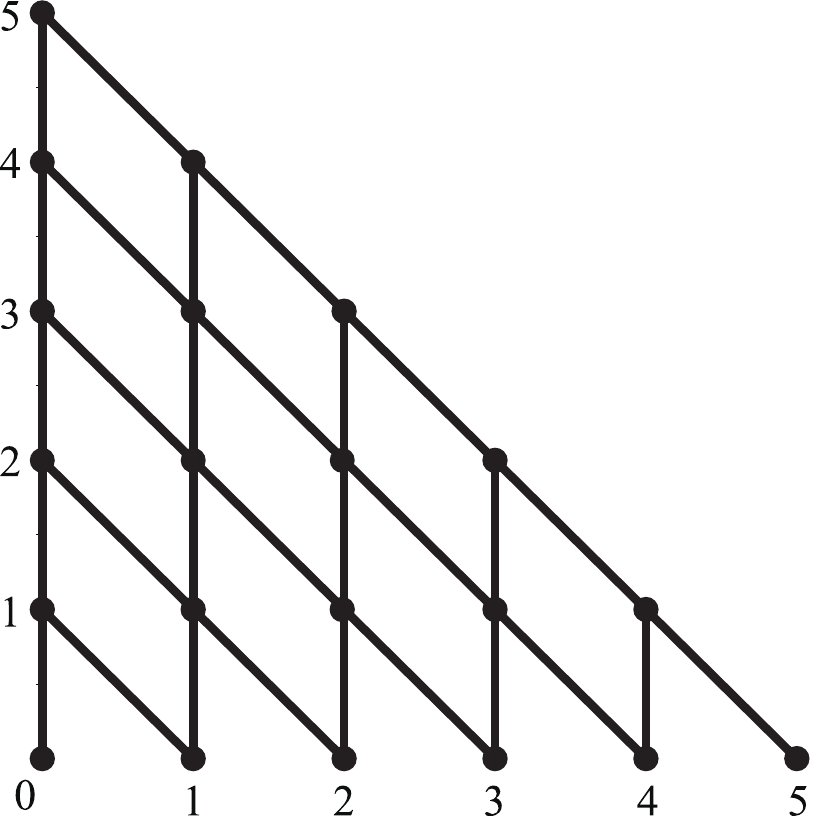}
        \end{center}
      \end{minipage}
\caption{The projected graphs associated with $\alpha A_{(1,0)}+\beta A_{(0,1)}$ (left) and $A_{(0,1)}$ (right) when $N=5$. On the graph associated with $A_{(0,1)}$, PST from (0,0) to $(N,0)$ occurs at some time $T$.}
\label{fig:a01graph}
\end{figure}

It should be noted here that PST also occurs on the graph $G_{0,1}$, whose projected lattice is of the shape given in Fig \ref{fig:a01graph} and that the graph coincides with one in \cite{Feder} when $N=2,3$. On all these graphs, PST occurs from $(0,0)$ to the farthest point $(N,0)$, which is desirable for quantum communication. 

\begin{figure}[htbp]
  \begin{center}
    \begin{tabular}{c}
      % 1
      \begin{minipage}{0.33\hsize}
        \begin{center}
          \includegraphics[width=3cm]{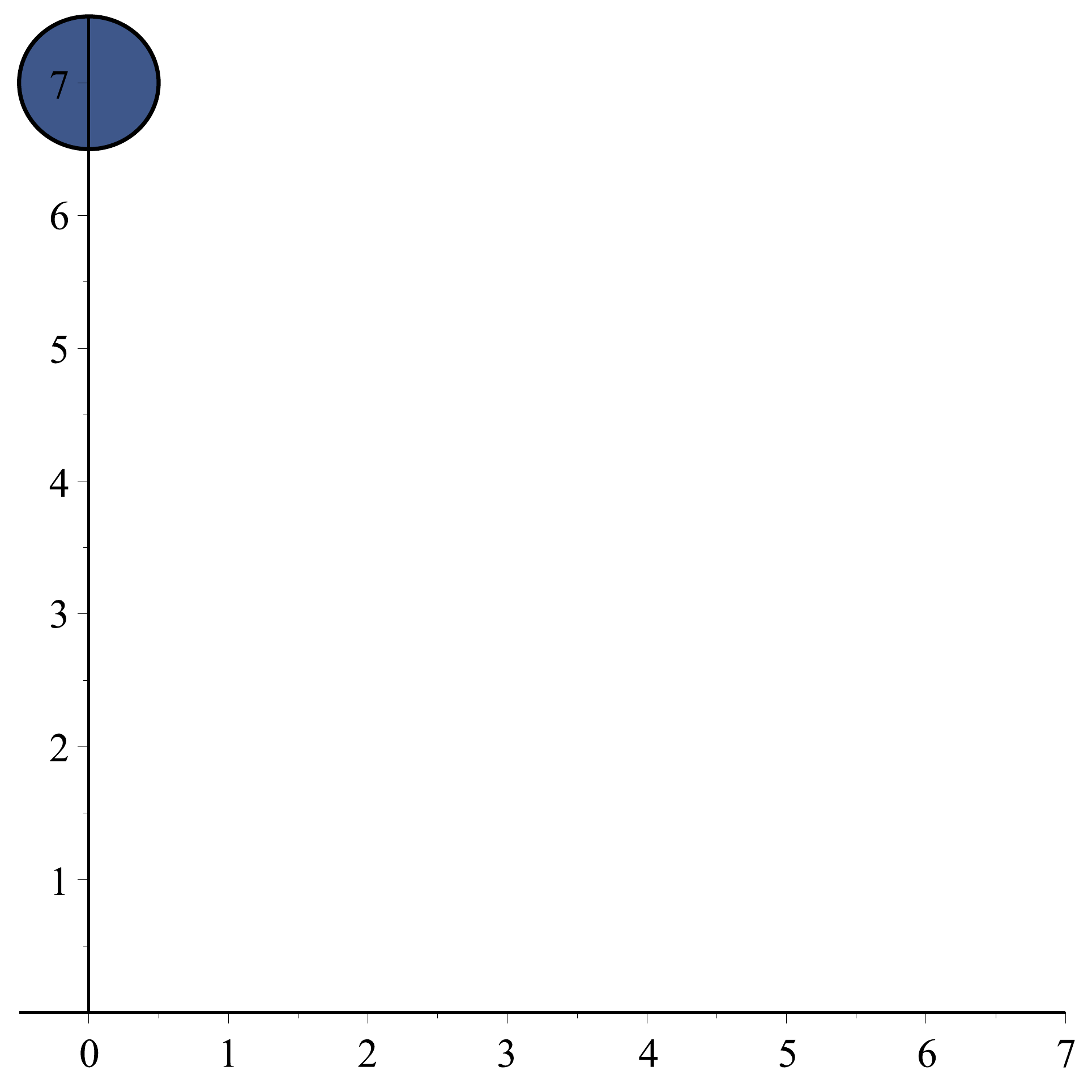}
         \hspace{1.6cm} $t=0$
        \end{center}
      \end{minipage}

      % 2
      \begin{minipage}{0.33\hsize}
        \begin{center}
          \includegraphics[width=3cm]{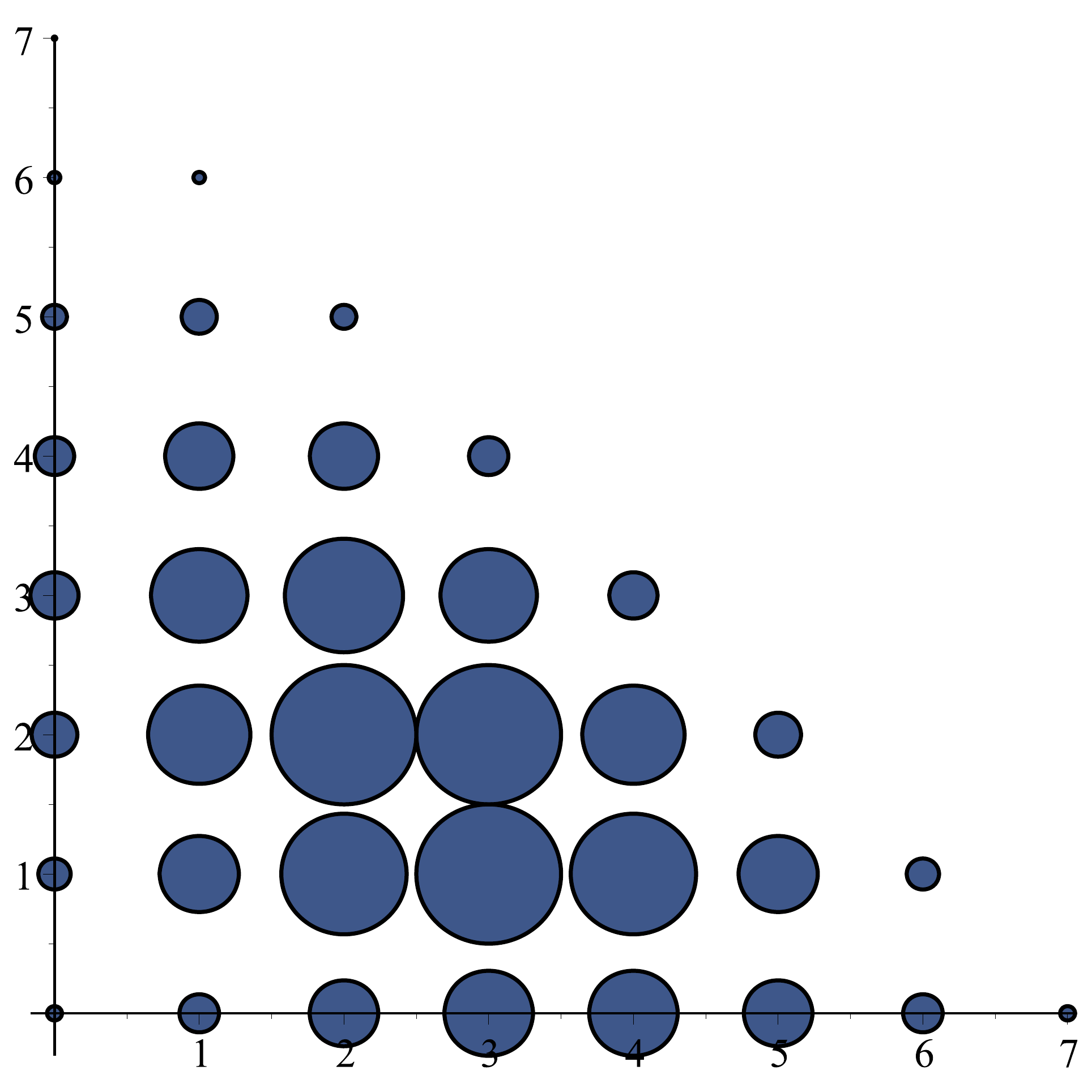}
                   \hspace{1.6cm} $t=\frac{\pi }{6}$
        \end{center}
      \end{minipage}

      % 3
      \begin{minipage}{0.33\hsize}
        \begin{center}
          \includegraphics[width=3cm]{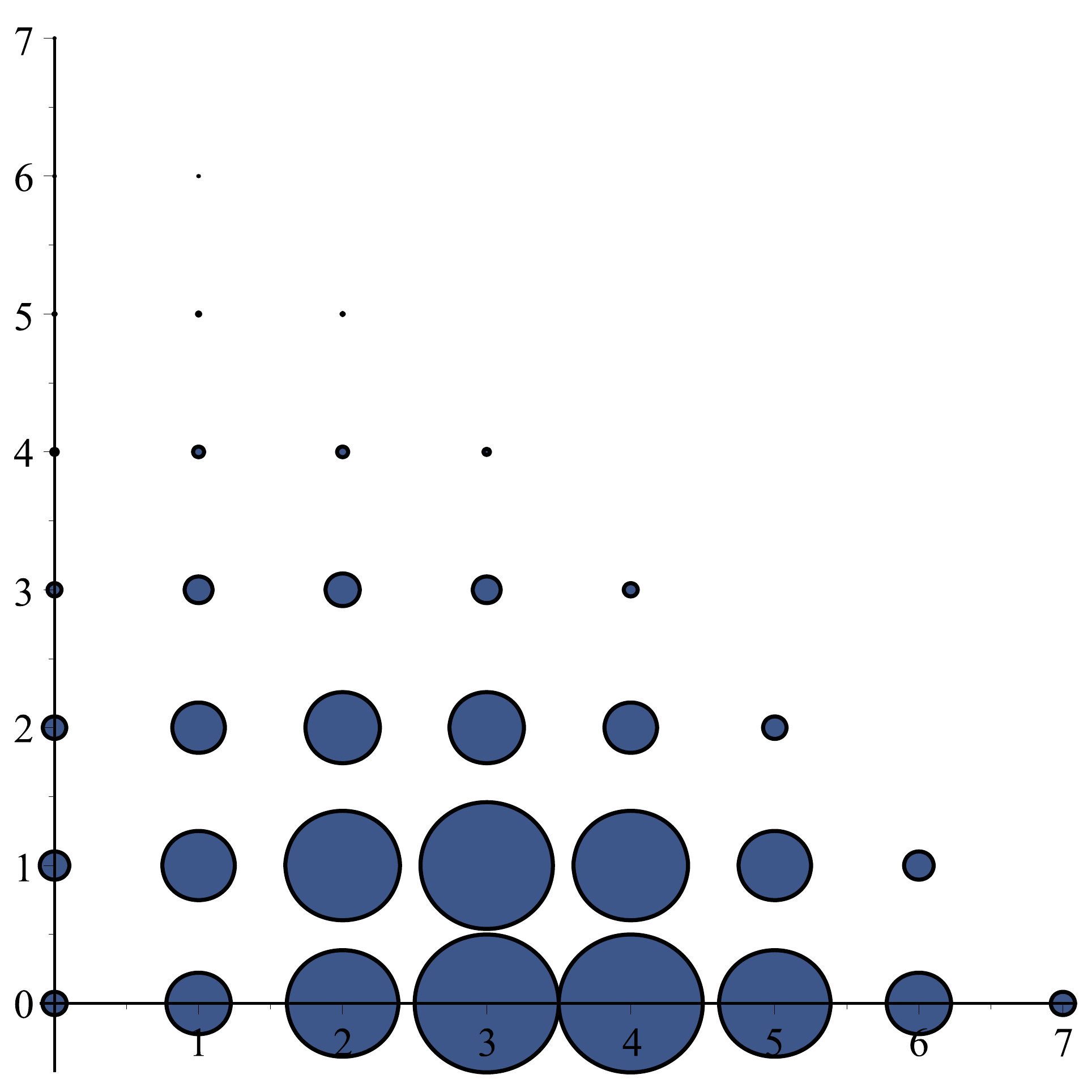}
                   \hspace{1.6cm} $t=\frac{\pi}{5}$
        \end{center}
      \end{minipage}
    \end{tabular}
        \begin{tabular}{c}

      % 1
      \begin{minipage}{0.33\hsize}
        \begin{center}
          \includegraphics[width=3cm]{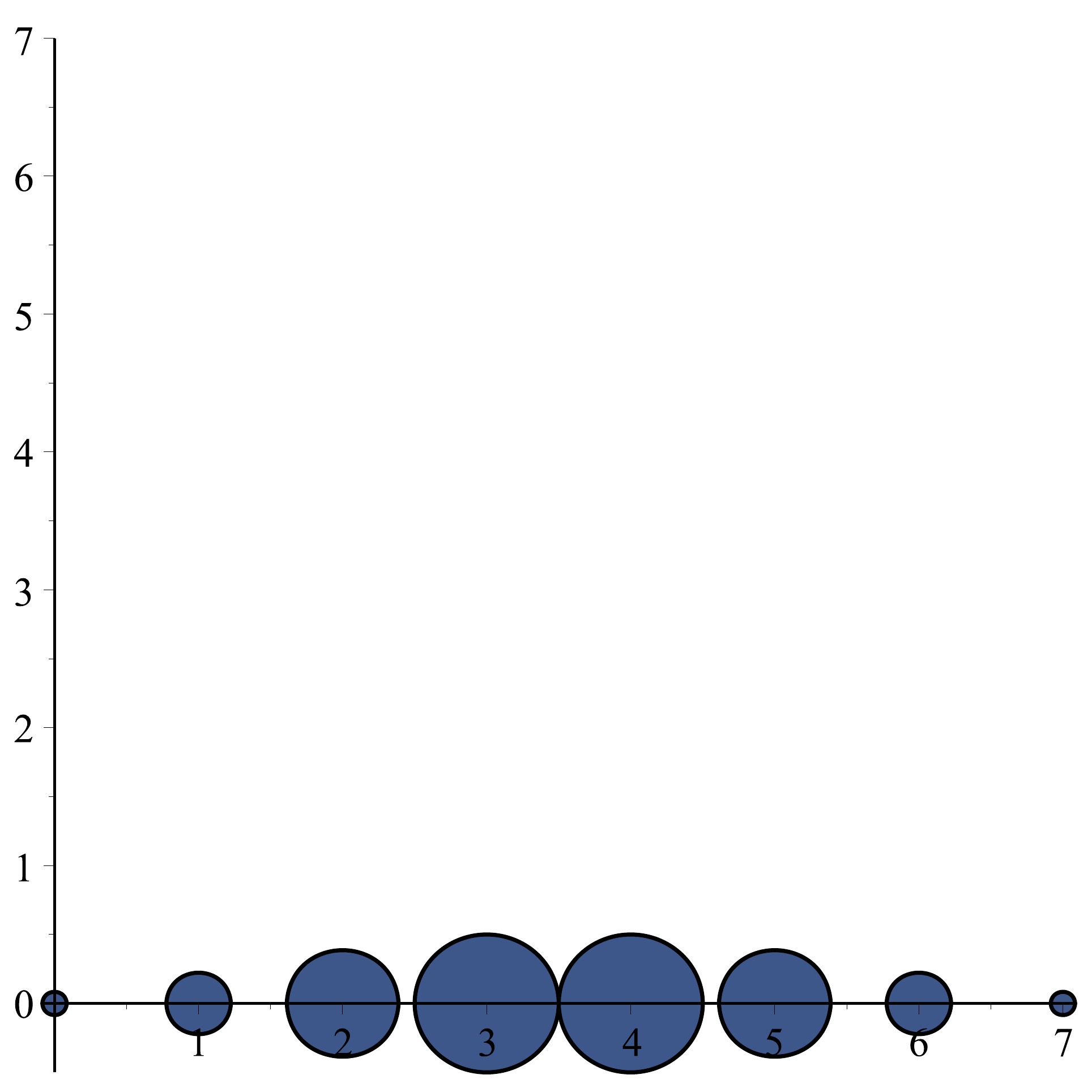}
                   \hspace{1.6cm} $t=\frac{\pi}{4}$
        \end{center}
      \end{minipage}

      % 2
      \begin{minipage}{0.33\hsize}
        \begin{center}
          \includegraphics[width=3cm]{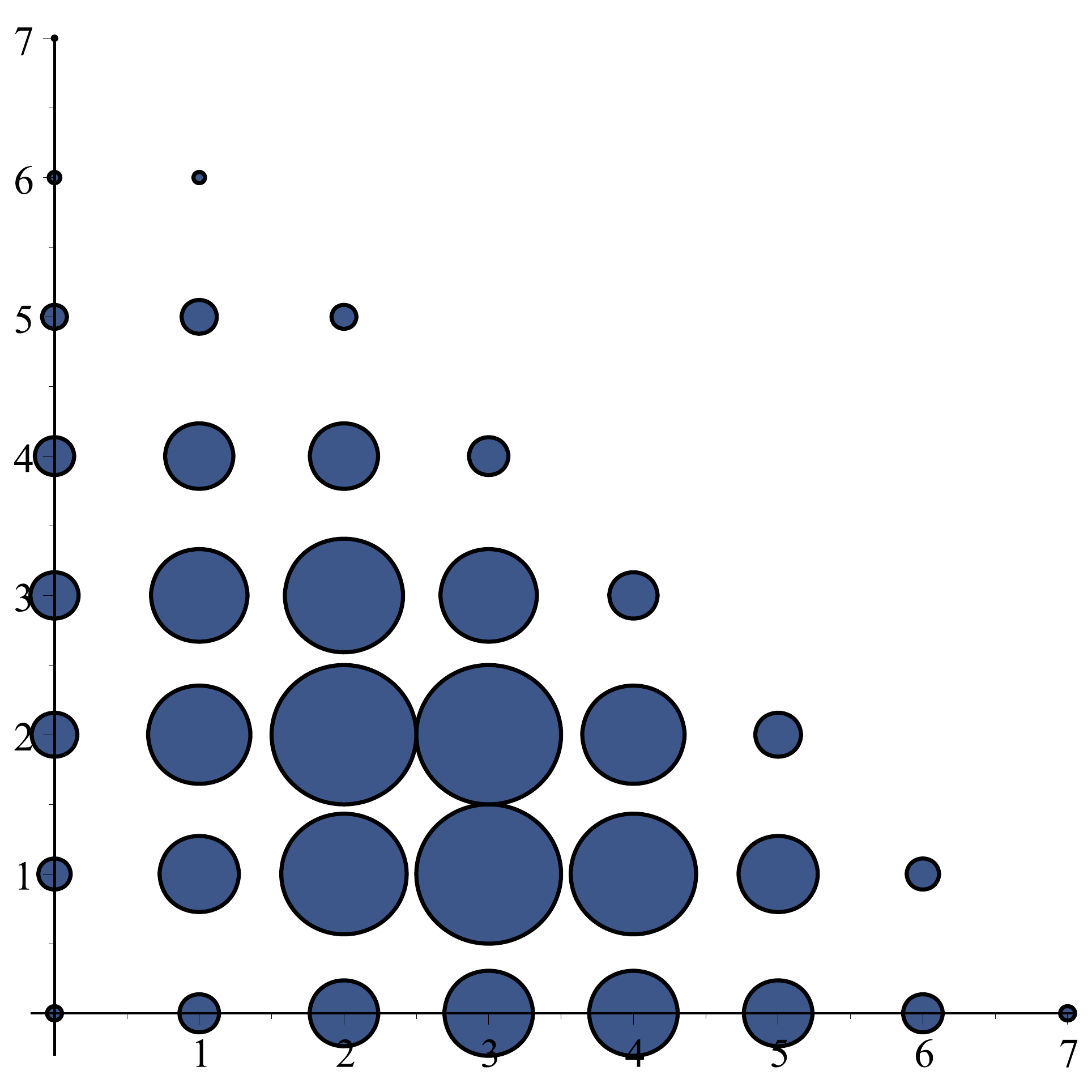}
                   \hspace{1.6cm} $t=\frac{\pi }{3}$
        \end{center}
      \end{minipage}

      % 3
      \begin{minipage}{0.33\hsize}
        \begin{center}
          \includegraphics[width=3cm]{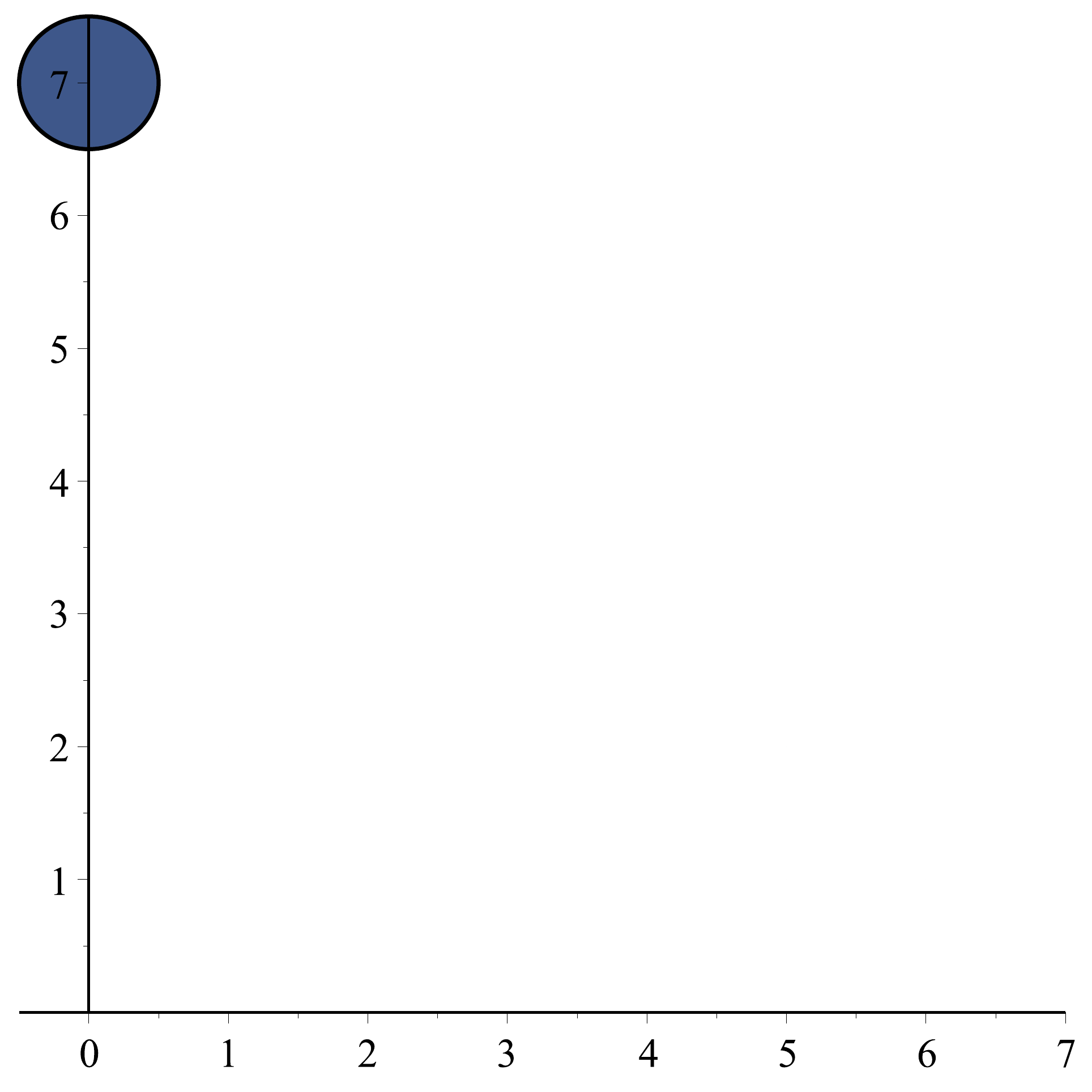}
                   \hspace{1.6cm} $t=\frac{\pi}{2}$
        \end{center}
      \end{minipage}
    \end{tabular}
    \caption{The transition amplitude $|g_{i,j}(t)|$ for $\sqrt{2}A_{(1,0)}+A_{(0,1)}$ when $N=7$. The areas of the circles
are proportional to $|g_{(i,j)}(t)|$ at the given lattice point $(i,j)$. FR on the set of sites $i=0,1,\cdots ,N$ and $j=0$ occurs at $\frac{\pi}{4}$.}
    \label{fig:su3}
  \end{center}
\end{figure}

It was remarked that when $\beta=\frac{\alpha}{\sqrt{2}}$, the hopping terms in the Hamiltonian \eqref{hamiltonian-action} are symmetric under rotations by $\frac{2}{3}\pi$.
The spin network then identifies with the weight lattice of the fully symmetrized tensor product of the fundamental representation of $SU(3)$.
That the bivariate Krawtchouk polynomials have an algebraic interpretation based upon $SU(3)$ has been established in \cite{Iliev} (see also \cite{Genest2}).
For this specific choice of parameters ($\beta=\frac{\alpha}{\sqrt{2}}$), interestingly it is found that there is FR between the site $(0,N)$ and the lattice points $(i,0)~~(i=0,1,\ldots ,N)$. Indeed, for the transition amplitude 
\[
g_{(i,j)}(t)=\left( e_{i,j}| e^{-itH} | e_{0,N}\right),
\]
there exists some time $T$ such that
\[
\sum_{i=0}^N |g_{(i,0)}(T)|^2=1.
\]
This is illustrated in Fig. \ref{fig:su3}.

\section{Concluding Remarks}

This paper has established the connection between quantum walks on graphs of the ordered Hamming scheme of depth 2 and the single excitation dynamics of certain two-dimensional lattices of triangular shape. This relation has featured the bivariate Krawtchouk polynomials of the Tratnik type that appear as eigenvalue matrices of the scheme and whose recurrence coefficients provide the couplings and Zeeman terms.
We have focused on Hamiltonians $\alpha  A_{(1,0)}+\beta A_{(0,1)}$ given by weighted combinations of the adjacency matrices of the two graphs associated to the shapes $(1,0)$ and $(0,1)$. Remarkably, when $\frac{\alpha }{\beta }$ is some rational number, we have observed that PST takes place between the sites $(0,0)$ and $(N,0)$ of the lattice at time $t=\frac{\pi }{2}$ after mixing on the whole two-dimensional lattice. In some examples, it has also been found that fractional revival occurs at $t=\frac{\pi}{4}$ at each of the sites of one side only of the lattice.\par 
It should be stressed that the spin lattice that has been found here differs from the one discussed in \cite{Miki1} which is based on the more general Krawtchouk polynomials of Griffiths \cite{Diaconis1,Genest2,Grunbaum1,Hoare1}. The question of determining the graph to which the model in \cite{Miki1} lifts thus remains.
The results presented here enrich the catalog of pairings between quantum walks on graphs and spin models in the context of PST. It is likely that PST could be preserved in the higher spin simplices related to graphs of the ordered Hamming scheme of depth $r$ where the multivariate Krawtchouk polynomials will intervene. It would prove interesting if such coherent transport could be realized in photonic lattices (see for instance \cite{Christandl1,Nikolopoulos1}). Finally, we would like to examine if the peculiar transport properties of the spin lattices could be of use in the design of certain algorithms.

\section*{Acknowledgements}
The authors would like to thank Matthias Christandl for asking about lifts to graphs
of coherent transport on spin lattices. They are grateful to Paul Terwilliger for bringing
references \cite{Barg} and \cite{Mizukawa} to their attention. They also thank Ryo Sato and Kengo Miura for discussions. The insightful inputs from Kareljan Schoutens and William Martin has also been much appreciated. LV wishes to acknowledge the hospitality of Kyoto University where most of this research was carried out. 
The research of ST is supported by JSPS KAKENHI (Grant Numbers 16K13761) and that of LV by a discovery grant of the Natural Sciences and Engineering Research Council (NSERC) of Canada.

% TODO:
% Provide your bibliography here. You have two options:

% FIRST OPTION - write your entries here directly, following the example below, including Author(s), Title, Journal Ref. with year in parentheses at the end, followed by the DOI number.
%\begin{thebibliography}{99}
%\bibitem{1931_Bethe_ZP_71} H. A. Bethe, {\it Zur Theorie der Metalle. i. Eigenwerte und Eigenfunktionen der linearen Atomkette}, Zeit. f{\"u}r Phys. {\bf 71}, 205 (1931), \doi{10.1007\%2FBF01341708}.
%\bibitem{arXiv:1108.2700} P. Ginsparg, {\it It was twenty years ago today... }, \url{http://arxiv.org/abs/1108.2700}.
%\end{thebibliography}

 \bibliographystyle{SciPost_bibstyle} % Include this style file here only if you are not using our template

\begin{thebibliography}{plain}
% A
\bibitem{Albanese1} C. Albanese, M. Christandl, N. Datta and A. Ekert, {\it Mirror inversion of quantum states in linear registers}, Phys. Rev. Lett. \textbf{93}, 230502 (2004), \doi{10.1103/PhysRevLett.93.230502}
% Ban
\bibitem{Banchi1} L. Banchi, E. Compagno and S. Bose, {\it Perfect wave-packet splitting and reconstruction in a one-dimensional lattice}, Phys. Rev. A \textbf{91}, 052323 (2015), \doi{10.1103/PhysRevA.91.052323}
% Bar
\bibitem{Barg} A. Barg and P. Purkayastha, {\it Bounds on ordered codes and orthogonal arrays}, Moscow Math. Journal \textbf{9}, 211--243 (2009) \doi{10.1109/ISIT.2007.4557247}
% Be
\bibitem{Bernard} P. Bernard, A. Chan, E. Loranger, C. Tamon and L. Vinet, {\it A graph with fractional revival} Phys. Lett. A \textbf{382}, 259--264 (2018) \doi{10.1016/j.physleta.2017.12.001}
%Bi
\bibitem{Bierbrauer} J. Bierbrauer, {\it A direct approach to linear programming bounds for codes and tms-nets}, Designs, Codes and Cryptography \textbf{42}, 127--143 (2007) \doi{10.1007/s10623-006-9025-6}
% Bose
\bibitem{Bose1} S. Bose, {\it Quantum communication through spin chain dynamics: an introductory review}, Contemp. Phys. \textbf{48}, 13--30 (2007) \doi{10.1080/00107510701342313}
% Boss
\bibitem{Bosse1} \'{E}.O. Bosse and L. Vinet, {\it Coherent transport in photonic lattices: a survey of recent analytic results}, SIGMA \textbf{13}, 074 (2017) \doi{10.3842/SIGMA.2017.074}
% Br
\bibitem{Brouwer1} A.E. Brouwer, A.M. Cohen and A. Neumaier, {\it Distance-Regular Graphs}, Springer (1989)
% ChiFG
\bibitem{Childs2} A. Childs, E. Farhi and S. Gutmann, {\it An example of the difference between quantum and classical random walks} Quant. Inf. Process. \textbf{1}, 35--43 (2002) \doi{10.1023/A:1019609420309}
% ChiG
\bibitem{Childs1} A. Childs and J. Goldstone, {\it Spatial search by quantum walk}, Phys. Rev. A \textbf{70}, 022314 (2004) \doi{10.1103/PhysRevA.70.022314}
% Chr
\bibitem{Christandl1} M. Christandl, N. Datta, T.C. Dorlas, A. Ekert, A. Kay and A.J. Landahl, {\it Perfect transfer of arbitrary states in quantum spin networks}, Phys. Rev. A \textbf{71}, 032312 (2005) \doi{10.1103/PhysRevA.71.032312}
% D
\bibitem{Diaconis1} P. Diaconis and R. Griffiths, {\it An introduction to multivariate Krawtchouk polynomials and their applications}, J. Stat. Plan. Infer. \textbf{154}, 39--53 (2014) \doi{10.1016/j.jspi.2014.02.004}
% FG
\bibitem{Farhi1} E. Farhi and S. Gutmann, {\it Quantum computation and decision trees}, Phys. Rev. A \textbf{58}, 915 (1998) \doi{10.1103/PhysRevA.58.915}
%Fe
\bibitem{Feder} D.L. Feder, {\it Perfect quantum state transfer with spinor bosons on weighted graphs}, Phys. Rev. Lett. \textbf{97}, 180502 (2006) \doi{10.1103/PhysRevLett.97.180502}
%GeVZ
\bibitem{Genest2} V. Genest, L. Vinet and A. Zhedanov, {\it The multivariate Krawtchouk polynomials as matrix elements of the rotation group representations on oscillator states}, J. Phys. A: Math. Theor. \textbf{46}, 505203 (2013) \doi{10.1088/1751-8113/46/50/505203}
\bibitem{Genest1} V. Genest, L. Vinet and A. Zhedanov, {\it Quantum spin chains with fractional revival}, Ann. Phys. \textbf{371}, 348--367 (2016) \doi{10.1016/j.aop.2016.05.009}
% Go
\bibitem{Godsil1} C. Godsil, {\it State transfer on graphs}, Disc. Math. \textbf{312}, 123--147 (2012) \doi{10.1016/j.disc.2011.06.032}
% Gr
\bibitem{Grunbaum1} F.A. Gr\"{u}nbaum and M. Rahman, {\it On a family of 2-variable orthogonal Krawtchouk polynomials}, SIGMA \textbf{6}, 090 (2010) \doi{10.3842/SIGMA.2010.090
}
% H
\bibitem{Hoare1} M.R. Hoare and M. Rahman, {\it A probabilistic origin for a new class of bivariate polynomials}, SIGMA \textbf{4}, 089 (2008) \doi{10.3842/SIGMA.2008.089}
% K
\bibitem{Iliev} P. Iliev and P. Terwilliger, {\it The Rahman polynomials and the Lie algebras $sl_3(\mathbb{C})$}, Trans. Amer. Math. Soc. \textbf{364}, 4225--4238 (2012) \doi{10.1090/S0002-9947-2012-05495-X}
\bibitem{Kay1} A. Kay, {\it Perfect, efficient, state transfer and its applications as a constructive tool}, Int. J. Quant. Inf. \textbf{8}, 641--676 (2010) \doi{10.1142/S0219749910006514}
\bibitem{Kendon1} V. Kendon and C. Tamon, {\it Perfect state transfer in quantum walks on graphs}, J. Comput. Theor. Nanosci. \textbf{8}, 422--433 (2011) \doi{10.1166/jctn.2011.1706}
% Ma
\bibitem{MartinStinson}
W.J. Martin and D.R. Stinson, {\it Association Schemes for Ordered Orthogonal Arrays and $(T, M, S)$-Nets},
Canad. J. Math. \textbf{51}, 326--346 (1999) \doi{10.4153/CJM-1999-017-5}
% Mik-M
\bibitem{Miki2} H. Miki and K. Miura, {\it 3-dimensional solvable XX spin lattice Hamiltonian derived from 3-variable Krawtchouk polynomials}, JSIAM Lett. \textbf{8}, 41--44 (2016) \doi{10.14495/jsiaml.8.41}
% Mik-TVZ
\bibitem{Miki1} H. Miki, S. Tsujimoto, L. Vinet and A. Zhedanov, {\it Quantum state transfer in a two-dimensional regular spin lattice of triangular shape}, Phys. Rev. A \textbf{85}, 062306 (2012) \doi{10.1103/PhysRevA.85.062306}
% Miz
\bibitem{Mizukawa} H. Mizukawa and H. Tanaka, {\it $(n+1,m+1)$-hypergeometric functions associated to character algebras}, Proc. Amer. Mathe. Soc. \textbf{132}, 2613-2618 (2004) \doi{10.1090/S0002-9939-04-07399-X}
% N
\bibitem{Nikolopoulos1} G.M. Nikolopoulos and I. Jex, {\it Quantum state transfer and network engineering}, Springer (2014)
% P
\bibitem{Post1} S. Post, {\it Quantum perfect state transfer in a 2D lattice}, Acta Appl. Math. \textbf{135}, 209--224 (2014) \doi{10.1007/s10440-014-9953-5}
% S
\bibitem{Stanton1} D. Stanton. {\it Orthogonal polynomials and combinatorics}, In \emph{Special Functions 2000: Current perspective and future directions}, J. Boustoz, M.E.H. Ismail, S. Suslov (eds.), NATO science series, \textbf{30}, 389--409,
Springer (2001) \doi{10.1007/978-94-010-0818-1_15}
% T
\bibitem{Tratnik1} M.V. Tratnik, {\it Some multivariable orthogonal polynomials of the Askey tableau-discrete families}, J. Math. Phys. \textbf{32}, 2337–-2342 (1991) \doi{10.1063/1.529158}

\end{thebibliography}

\nolinenumbers

\end{document}